\documentclass[dvipsnames]{article}
\usepackage[symbol]{footmisc}
\usepackage{rotating}
\usepackage{lineno}

\usepackage[a4paper,margin=1.1in,footskip=0.25in]{geometry}

\usepackage[utf8]{inputenc}
\usepackage[english]{babel}
\usepackage{csquotes}

\usepackage{titling}

\usepackage{graphicx}
\usepackage[usenames,dvipsnames,svgnames,table]{xcolor}

\usepackage{amssymb}
\usepackage{amsmath}
\usepackage{amsfonts}
\usepackage{algorithm}
\usepackage{algpseudocode}

\usepackage[backend=biber, style=nature, doi=false, maxbibnames=3]{biblatex}
\usepackage{xr}
\makeatletter
\newcommand*{\addFileDependency}[1]{
\typeout{(#1)}
\@addtofilelist{#1}
\IfFileExists{#1}{}{\typeout{No file #1.}}
}\makeatother

\usepackage{url}
\usepackage[hyperindex,colorlinks,hyperfootnotes = false,]{hyperref}

\usepackage{amsthm}

\title{Patterns of co-occurrent skills in UK job adverts \\
}

\makeatletter
\newcommand{\printfnsymbol}[1]{
  \textsuperscript{\@fnsymbol{#1}}
}
\makeatother

\author
{Zhaolu Liu$^{1}$, Jonathan M.\ Clarke$^{1}$, Bertha Rohenkohl$^{2}$ and Mauricio Barahona$^{1\ast}$
\\
\normalsize{$^{1}$Department of Mathematics, Imperial College London, London, United Kingdom}
\\
\normalsize{$^{2}$Institute for the Future of Work, London, United Kingdom}
\\
\normalsize{$^\ast$To whom correspondence should be addressed; E-mail:  m.barahona@imperial.ac.uk}
}

\date{} 

\begin{document}

\maketitle

\begin{abstract}
 \noindent A job usually involves the application of several complementary or synergistic skills to perform its required tasks. Such relationships are implicitly recognised by employers in the skills they demand when recruiting new employees. Here we construct a skills network based on their co-occurrence in a national level data set of 65 million job postings from the UK spanning 2016 to 2022. We then apply multiscale graph-based community detection to obtain data-driven skill clusters at different levels of resolution that reveal a modular structure across scales. Skill clusters display diverse levels of demand and occupy varying roles within the skills network: some have broad reach across the network (high closeness centrality) while others have higher levels of within-cluster containment, yet with high interconnection across clusters and no skill silos.  
 The skill clusters also display varying levels of semantic similarity, highlighting the difference between co-occurrence in adverts and intrinsic thematic consistency.  
 Clear geographic variation is evident in the demand for each skill cluster across the UK, broadly reflecting the industrial characteristics of each region, e.g., London appears as an outlier as an international hub for finance, education and business.  
 Comparison of data from 2016 and 2022 reveals employers are demanding a broader range of skills over time, with more adverts featuring skills spanning different clusters. We also show that our data-driven clusters differ from expert-authored categorisations of skills, indicating that important relationships between skills are not captured by expert assessment alone.
\end{abstract}

\renewcommand{\abstractname}{Author Summary}
\begin{abstract}
\noindent Jobs often require employees to apply a wide range of skills in their work. Understanding how these skills relate to one another is important to provide insight into how employees may be more or less able to carry out their jobs or find other jobs, as well as to track how occupations change over time, for instance when new technologies are introduced. In this study we use a large dataset of 65 million job adverts between 2016 and 2022 across the whole of the UK to examine the patterns of skills required together by employers. We find clusters of skills that appear together in adverts often, but these clusters do not always agree with how experts group skills based on competencies or qualifications. Overall, we find a strong co-requirement of varied skills by employers, with less interconnection for some technical skills, such as in cybersecurity. Which skill clusters are in demand varies significantly across the UK, with London standing out as an international hub for finance, education and business. Over time, skills in the UK labour market have become more interconnected, reflecting employers expecting workers to possess a more diverse range of skills to do their jobs.  
\end{abstract}

\section*{Introduction}
The association between a job and the skills it requires is complex. Often, jobs require 
a range of skills; some highly specialised for a role, others more generic and common to many occupations~\cite{alabdulkareem_unpacking_2018}. When filling a vacancy, employers may thus place emphasis on particular skills over others, or on a combination of skills.
With the emergence and adoption of new technologies spanning industries and occupations across the labour market, there is an increased need to study the complex and evolving skills requirements of employers.
%

 Traditionally, skills have been classified from the perspective of the employee, with a focus on educational history, qualifications and other competencies, many of which do not map neatly to the skills that employers require for a role~\cite{autor_growth_2013}. 
On the other hand, economists study the change in skills requirements of an economy using changes in the occupational composition of the economy as a proxy~\cite{autor_skill_2003,deming_skill_2018}, 
although this strategy cannot easily account for changes that occur within occupations
or commonalities in skill demand between occupations within specific industries or locations~\cite{deming_skill_2018}. 
Historically, labour markets have adapted to technological advancements with new roles being created; some jobs being displaced by automation; and others being complemented and augmented by new technologies~\cite{autor_why_2015}. 
With the current wave of technological progress, particularly the rapid advancements in AI and the emergence of new skills related to the creation, adaptation and use of automation technologies~\cite{cep, acemoglu_tasks_2022, acemoglu_autor_hazell}, 
there is a renewed emphasis on the role of combination of skills for firms and workers ~\cite{rohenkohl_what_2023,hayton_organisational_2023} to succeed in rapidly evolving modern labour markets~\cite{acemoglu_skills_2011, brynjolfsson_artificial_2017, acemoglu_tasks_2022}.

 The emergence of large job postings data sets in recent years has opened the opportunity to study in an agnostic, data-driven manner different aspects of the relationships between skills, jobs and the labour market.  For instance, recent work has examined which combinations of skills are demanded by employers for specific roles, and how skills may relate to each other in the labour market~\cite{stephany_what_2024, deming_skill_2018,vassilev_whats_nodate, cammeraat_burning_2021}. 
Such data sets have also been shown to capture vacancies of geographic regions and occupations~\cite{mahoney-nair_review_2021, cammeraat_burning_2021,bassier_vacancy_2023}, and to reflect changes in labour demand, as exemplified by their use by the UK Office for National Statistics (ONS) and the Organisation for Economic Co-operation and Development (OECD)~\cite{ons_adzuna,vassilev_whats_nodate, oecd_skills}, among others.  
Job postings data sets  
have also revealed the pay premium derived through possessing specific skills, and to examine how labour markets respond to the emergence of new technologies~\cite{acemoglu_autor_hazell, cao_technological_2022,  deming_earnings_2020}. 
Given that most jobs, and consequently job adverts, require several skills~\cite{vassilev_whats_nodate, stephany_what_2024, cep}, the study of 
modern labour markets must consider not only the prevalence of individual skills, but also their complementarity and synergy. 

The focus on relationships between skills lends itself naturally to network analysis methods, in the spirit of research in economic complexity,
%
where economic networks are built using empirical data that captures pairwise relationships between entities (countries, industries, firms) based on similarities of their economic profiles
~\cite{hidalgo_economic_2021,hidalgo_building_2009, balland_new_2022}. 
Data-driven similarities between skills have been previously based on occupation data. In the US, a skill occupation network was constructed from the O*NET database that maps skills to occupations based on a survey of US workers~\cite{onet}, and was   
%
shown to be predictive of worker transitions between occupations, while also revealing the competing roles of skills frictions and geographic frictions in determining job transitions~\cite{frank_network_2024,del_rio-chanona_occupational_2021}. Further, US cities with well connected skill occupation networks display greater economic resilience~\cite{moro_universal_2021}, and  
metropolitan areas with skills that have high network centrality are more productive and command higher salaries~\cite{waters_impacts_2022}. 

Here, we use a large data set of UK job postings collected by Adzuna Intelligence (65 million adverts spanning 2016 to 2022) to examine the relationships between skills across the UK labour market based on their co-occurrence in job adverts, as a reflection of the demand by employers~\cite{adzuna_link}.
To capture the skills landscape of the whole UK labour market, 
we use this national-level data set of job postings to create a \emph{skills co-occurrence network} by implementing a graph construction protocol, which employs both dimensionality reduction and a consistent geometric graph definition, to produce a sparsified skills co-occurrence network that captures the global and local geometry of the data. 
This skills network is analysed using  Markov stability (MS), a multiscale graph-based clustering technique to extract data-driven groupings of skills with consistent co-occurrence across a range of resolutions, from fine to coarse. 
These data-driven skills clusters show strong thematic coherence, although not strictly concomitant with standard expert-based categories.
Focusing on a medium resolution clustering, whereby the 3906 individual skills are grouped into 21 skill clusters, we use the centrality and containment of skill clusters to evaluate the extent to which skills are required alongside skills in the same or different skill clusters, and how such network properties  relate to estimated wages. Finally, we explore the variation in the demand for skills clusters  both in time (from 2016 to 2022) as well as geographically 
(across UK regions) to gain insights about temporal trends and regional economic differences. 
%



\section*{Results}

\subsection*{From job adverts to a skills network}

Our analysis is carried out on a curated and deduplicated data set containing 65 million job adverts posted in the UK collected weekly during
2016 (11 million adverts, 1.2 million/month over 9 months), 2018 (18 million, 1.5 million/month, 12 months), 2020 (16 million, 1.3 million/month, 12 months) and 2022 (20 million, 1.6 million/month, 12 months) for an average of 1.4 million adverts per month.
Each advert has its date of first posting, geographical location, and is linked to at least one skill out of the 3906 skills in the Lightcast Open Skills  taxonomy. Crucially, 99.6\% of adverts contain at least one skill. For a full description of the data set and preprocessing, see Methods.

The total mentions of skills in the data set is 634 million, i.e., each advert is linked to 9.4 skills on average.
This means that there is a rich source of information in the \textit{co-occurrence} of skills within each advert. As described in Methods, we summarise the patterns of co-occurrence by constructing a sparsified weighted undirected similarity graph, $\mathcal{G}$, where the skills are nodes and they are connected according to the similarity of their co-occurrence in job adverts. We then analyse the skills network $\mathcal{G}$ by computing several network properties and through multiscale community detection.
The pipeline of the analysis is summarised in the flowchart in Figure \ref{fig:flowchart} in Methods.

\subsection*{The centrality of skills in the network}
\begin{figure}[htb!]
    \centering
    \includegraphics[width=1\textwidth]{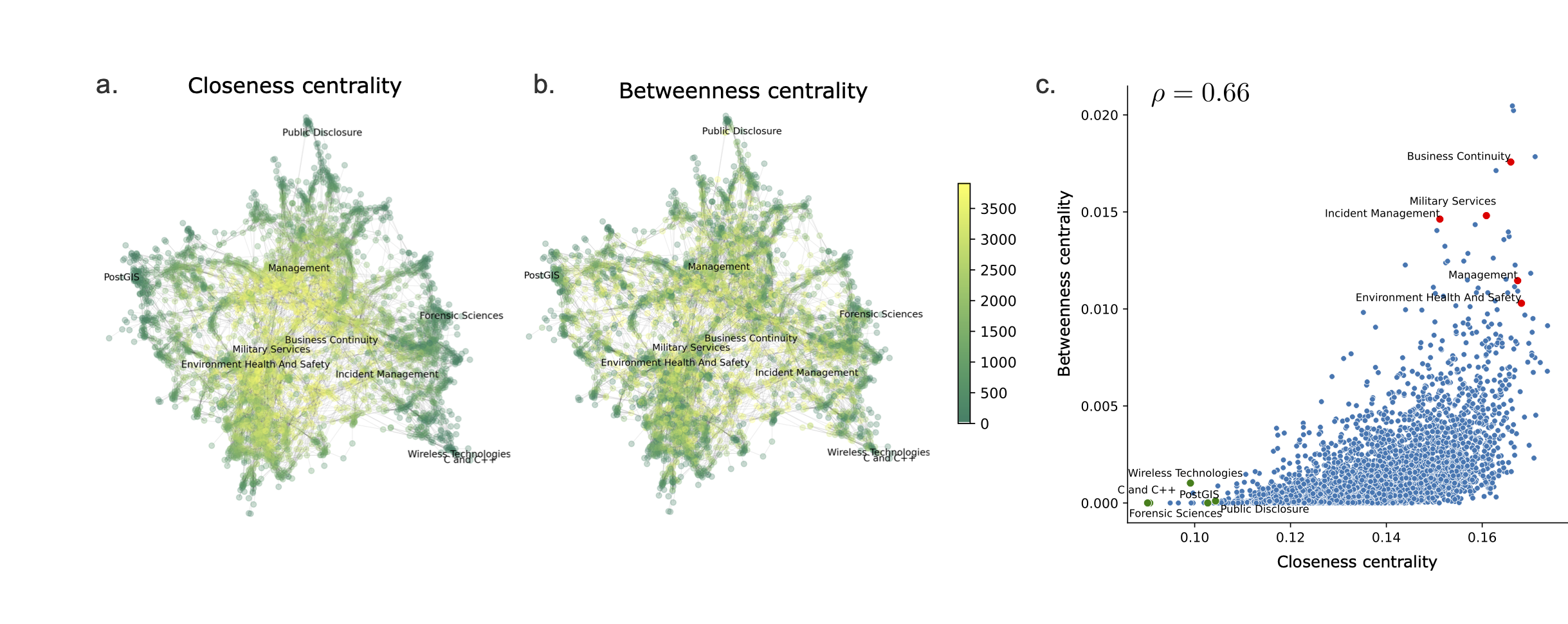}
    \caption{\textbf{Centrality in the skills network.} Skills network with nodes colored by (a) closeness centrality rank and (b) betweenness centrality rank. 
    Yellow indicates high centrality rank, while green indicates low centrality rank. (c) The scatter plot between both centralities shows moderate correlation between them. Some highly mentioned skills with high (red dots) and low (green dots) centrality are indicated. 
    }
    \label{fig:full_network_summary}   
\end{figure}

The skills network $\mathcal{G}$ can be studied using different tools from network science. A key concept in networks is node centrality~\cite{newman2018networks}, which measures the importance of nodes in the network. 
Two examples of network centrality measures are closeness and betweenness, both based on computing the shortest paths across all nodes in the graph~\cite{arnaudonScaledependentMeasureNetwork2020}. 
The closeness centrality of a node is the average length of the shortest paths between that node and all the other nodes in the graph; hence closeness measures how easy it is to reach all other parts of the network from that node. 
The betweenness centrality of a node counts the number of shortest paths (between any two nodes in the graph) that go through that node; hence betweenness measures how critical a node is to connect the different parts of the network. Figure~\ref{fig:full_network_summary} shows that these centralities are related, but weakly, in our skills network. 


%

Here, we use these measures of centrality to characterise the skills (nodes) according to the patterns of connectivity in the co-occurrence network~\cite{waters_impacts_2022,stephany_what_2024}. 
Skills with high closeness centrality constitute a common ground of skills shared across many different types of adverts, whereas skills with high betweenness centrality correspond to skills that bridge disparate groupings of skills that have less in common. 
Figure~\ref{fig:full_network_summary}a,b shows the skills network $\mathcal{G}$ with skills (nodes) coloured by their closeness and betweenness centrality, respectively. As expected, skills with high closeness centrality lie close to the core of the network, whereas skills with high betweenness centrality also appear as bridges from the more external parts of the network. 

Figure~\ref{fig:full_network_summary}c shows that closeness and betweenness centralities are correlated, indicating that the core of shared skills bridge far away groups of skills, but with some notable deviations. 
Relatively few skills have high betweenness centrality, and all that do also have high closeness centrality. Broadly, these skills appear to be relatively generic and relate to activities with higher levels of responsibility including `Management', `Business Continuity' and `Military Services'). These skills are both close to many skills in the network, but also connect skills that are otherwise poorly connected. 
Conversely, skills with the lowest betweenness and closeness centrality are largely specific, technical skills that relate to a small number of occupations or industrial sectors, including `Forensic Science', `Public Disclosure' and `Wireless Technologies'. Each of these skills are found at the periphery of the skills network and are poorly connected to the skills network as a whole. 

\subsection*{The multiscale structure of co-occurrence in the skills network}

\begin{figure}[htb!]
    \centering
    \includegraphics[width=0.80\textwidth]{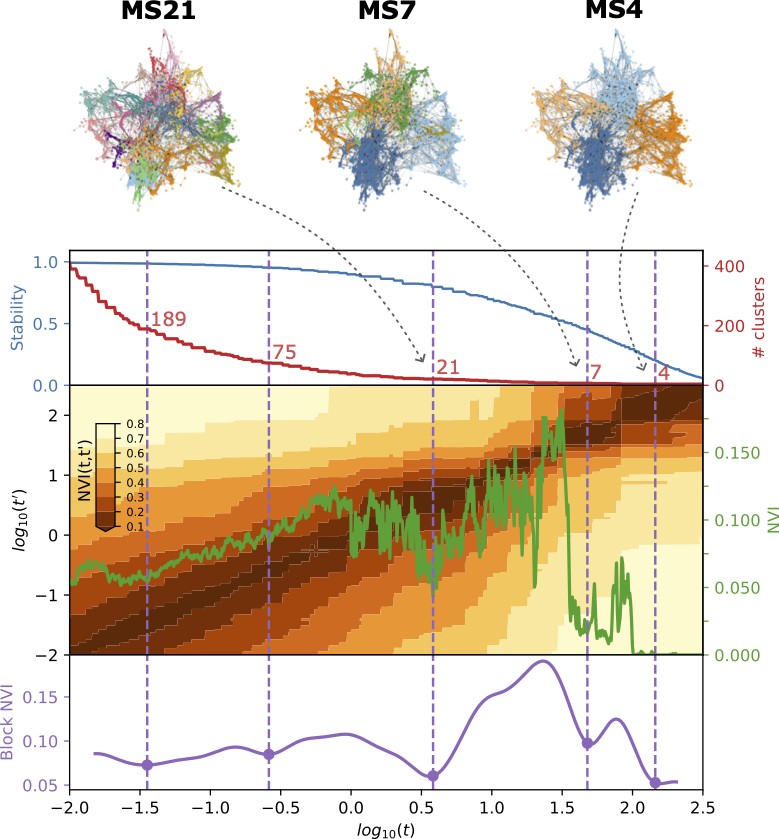}
    \caption{\textbf{Optimal clusterings from Markov Stability on the sparsified graph $\mathcal{G}$.} Five optimal highly robust partitions are identified based on minima of the Block NVI. For further details, see text,  Refs.~\cite{arnaudon2023pygenstability,lambiotte2014random,schaub2014structure,schindler2023multiscale}, and Appendix~\ref{app:MS}.
    } 
    \label{fig: markov_stability}   
\end{figure}

Next, we studied the skills network $\mathcal{G}$ using community detection to extract groups of skills that have similar patterns of co-occurrence in our data set. In this formulation, the skill clusters correspond to communities (i.e., subgraphs in the network) with strong similarities within the group. 
Here, we apply Markov Stability (MS)~\cite{
delvenne2010stability,schaub2012markov, delvenne2013stability}, an unsupervised multiscale graph clustering method, which reveals intrinsic, robust clusters of skills at different levels of resolution. The communities are obtained using a diffusion in the network biased by the likelihood of co-occurrence, thus leading to groups of skills that are consistently shared in job adverts.
Our computations are carried out using the Python package PyGenStability~\cite{arnaudon2023pygenstability}. 
For full details see Methods and Appendix~\ref{app:MS}.
 
\begin{figure}[htb!]
    \centering
    \includegraphics[width=1\textwidth]{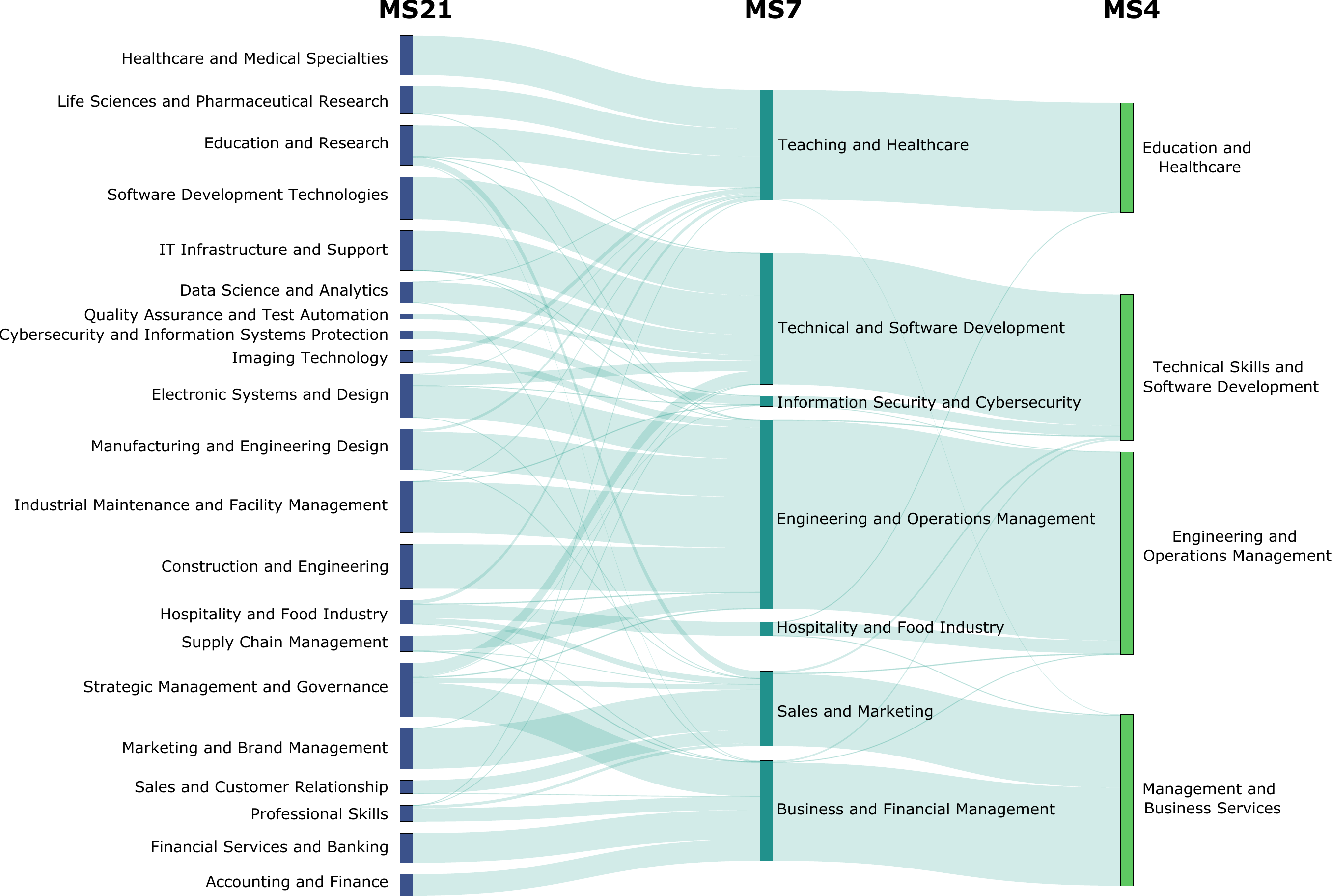}
     \caption{\textbf{Sankey diagram capturing the multiscale clustering of skills at different levels of resolution.} The quasi-hierarchical structure of skill co-occurrences is not imposed by the method but emerges naturally from the intrinsic co-occurrence patterns in the data. 
    } 
    \label{fig: sankey_allconfig}
\end{figure}

Figure~\ref{fig: markov_stability} summarises the Markov Stability analysis for our skills network. As signalled by minima of the block Normalised Variation of Information (NVI), we find five robust partitions of the 3906 skills into \textit{skill clusters} of different coarseness, from fine to coarse. Notably, as seen in the Sankey diagram (Figure~\ref{fig: sankey_allconfig}), 
the partitions have a strong quasi-hierarchical structure. This feature, which is not imposed \textit{a priori}  by our clustering method,  reveals an inherent consistency in how skill clusters relate to each other across levels of resolution: smaller clusters of skills that co-occur consistently get grouped into larger skill clusters with looser co-occurrence patterns. 

In this paper, we choose to examine in detail the  partition with maximal robustness corresponding to a medium level of resolution (MS21, 21 clusters). 
To aid interpretation of the obtained skill clusters, we developed an automated approach to create labels for the communities based on the properties subgraph and semantic summarisation via a large language model (see Methods).
In Appendix~\ref{sec:MS7}, we also include the full analysis of the coarser partition (MS7) into 7 skill clusters.

\subsection*{Clusters of co-occurrent skills from job adverts} 

As the main focus for our analysis, we consider the partition of the skills network into 21 data-driven skill clusters obtained using  Markov Stability (MS21). 
This medium resolution partition provides sufficient granularity in the clusters to usefully identify distinct skills communities, while representing a robust, stable partition of the skills network, as assessed by the low value of the block NVI (Fig.~\ref{fig: markov_stability}).

\subsubsection*{Characterisation of the skill clusters}

\begin{figure}[htb!]
    \centering
    \includegraphics[width=.92\textwidth]{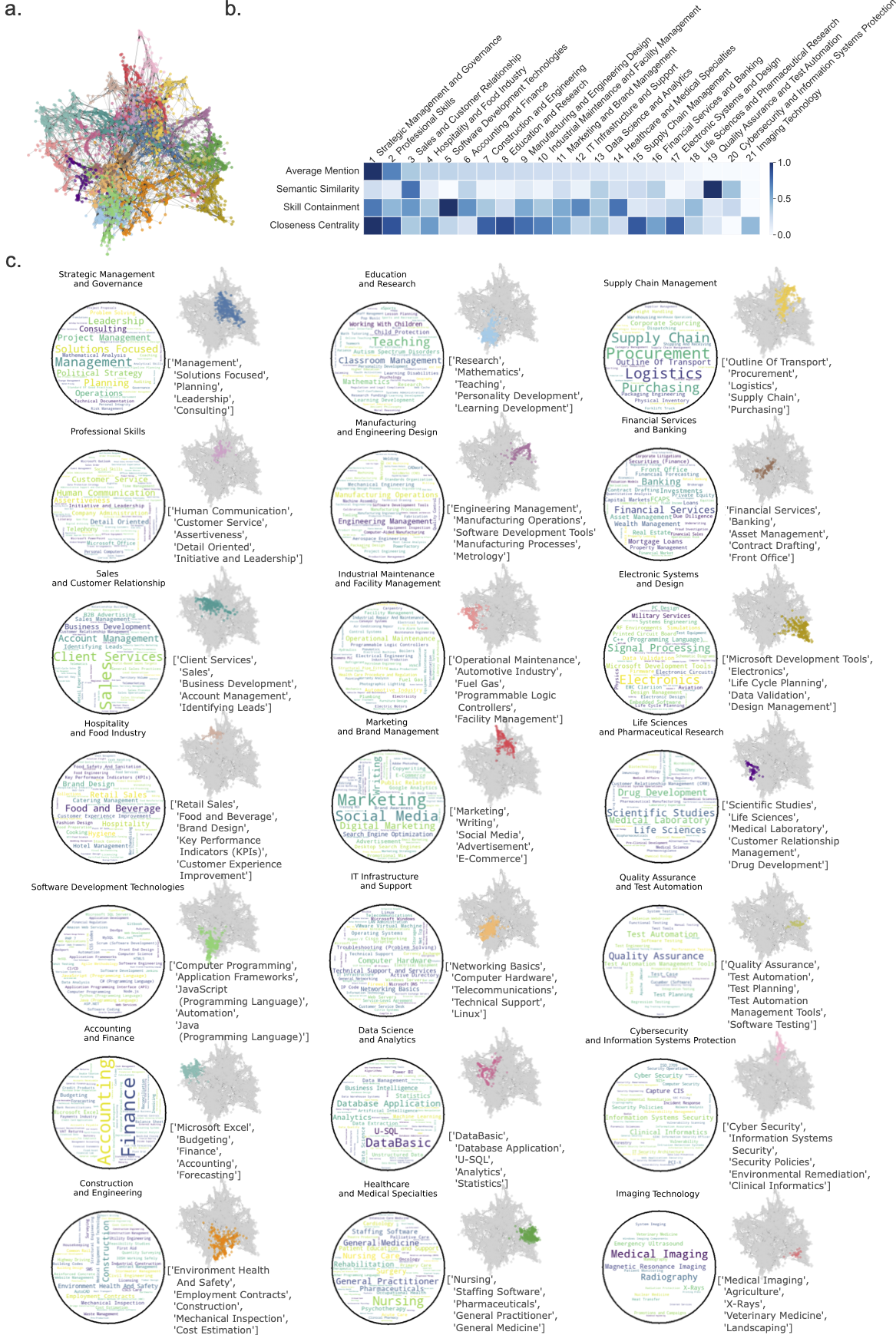}
    \caption{\textbf{Co-occurrence skill clusters (MS21)} (a) Skills network coloured according to the 21 skill clusters. (b) Summary heatmap of skill clusters properties. Each row is normalised by its maximum. (c) For each of the 21 clusters, word cloud where font size represents skill eigenvector centrality, and list of top 5 most frequent skills. 
    }
    \label{fig: MS21_fullinfo}
\end{figure}

Table~\ref{tab: descriptive_table} and Figure~\ref{fig: MS21_fullinfo} present a summary of the 21 skill clusters in MS21. 
The number of skills in each cluster 
varies from the largest  cluster `Strategic Management and Governance' (329 skills) to the smallest cluster `Quality Assurance and Test Automation' (31 skills). 
The labels for each cluster were generated automatically from the most central skills using Llama 2, as described above, and the clusters contain distinct groupings of skills with consistent thematic links, as shown in the word clouds (Figure~\ref{fig: MS21_fullinfo}c). 

\begin{table}[htb!]
\centering
\caption{Summary of properties of the medium resolution data-driven skill clusters (MS21).
}
\label{tab: descriptive_table}
\resizebox{.95\textwidth}{!}{%
\begin{tabular}{lllllllll}
\hline
 & Cluster & \begin{tabular}[c]{@{}l@{}}Number\\ of Skills\end{tabular} & \begin{tabular}[c]{@{}l@{}}Number \\ of Mentions\end{tabular} & \begin{tabular}[c]{@{}l@{}}Average\\ Mentions\end{tabular} & \begin{tabular}[c]{@{}l@{}}Semantic\\ Similarity\end{tabular} & \begin{tabular}[c]{@{}l@{}}Skill\\ Containment\end{tabular} & \begin{tabular}[c]{@{}l@{}}Closeness\\ Centrality\end{tabular} & \begin{tabular}[c]{@{}l@{}}Average\\ Salary\end{tabular} \\
\hline
1 & Strategic Management and Governance & 329 & 116654801 & 1.79 & 0.194 & 0.316 & 0.154 & 37190 \\
2 & Professional Skills & 100 & 79035037 & 1.22 & 0.167 & 0.229 & 0.149 & 25960 \\
3 & Sales and Customer Relationship & 81 & 43746343 & 0.67 & 0.318 & 0.276 & 0.130 & 30030 \\
4 & Hospitality and Food Industry & 150 & 35552886 & 0.55 & 0.167 & 0.198 & 0.138 & 24480 \\
5 & Software Development Technologies & 265 & 34966624 & 0.54 & 0.139 & 0.410 & 0.127 & 42270 \\
6 & Accounting and Finance & 129 & 33423443 & 0.51 & 0.212 & 0.285 & 0.123 & 32440 \\
7 & Construction and Engineering & 277 & 32640316 & 0.50 & 0.153 & 0.205 & 0.147 & 30520 \\
8 & Education and Research & 249 & 31458940 & 0.48 & 0.154 & 0.154 & 0.150 & 27020 \\
9 & Manufacturing and Engineering Design & 253 & 29459698 & 0.45 & 0.172 & 0.250 & 0.145 & 32610 \\
10 & Industrial Maintenance and Facility Management & 319 & 27501198 & 0.42 & 0.154 & 0.205 & 0.142 & 29710 \\
11 & Marketing and Brand Management & 249 & 25570594 & 0.39 & 0.192 & 0.248 & 0.136 & 31570 \\
12 & IT Infrastructure and Support & 250 & 23828152 & 0.36 & 0.158 & 0.296 & 0.127 & 35520 \\
13 & Data Science and Analytics & 126 & 20184523 & 0.31 & 0.220 & 0.200 & 0.131 & 37790 \\
14 & Healthcare and Medical Specialties & 241 & 19464635 & 0.30 & 0.187 & 0.322 & 0.125 & 30860 \\
15 & Supply Chain Management & 99 & 17582566 & 0.27 & 0.200 & 0.128 & 0.148 & 27520 \\
16 & Financial Services and Banking & 180 & 15899362 & 0.24 & 0.218 & 0.145 & 0.137 & 38220 \\
17 & Electronic Systems and Design & 277 & 13615819 & 0.21 & 0.146 & 0.175 & 0.147 & 36220 \\
18 & Life Sciences and Pharmaceutical Research & 176 & 6718425 & 0.10 & 0.169 & 0.223 & 0.128 & 36440 \\
19 & Quality Assurance and Test Automation & 31 & 2920005 & 0.04 & 0.381 & 0.129 & 0.122 & 38420 \\
20 & Cybersecurity and Information Systems Protection & 52 & 2468472 & 0.04 & 0.238 & 0.166 & 0.116 & 44630 \\
21 & Imaging Technology & 73 & 1311933 & 0.02 & 0.141 & 0.078 & 0.133 & 29140
\\
\hline
\end{tabular}%
}
\end{table}

%

\begin{figure}[htb!]
    \centering
\includegraphics[width=0.9\textwidth]{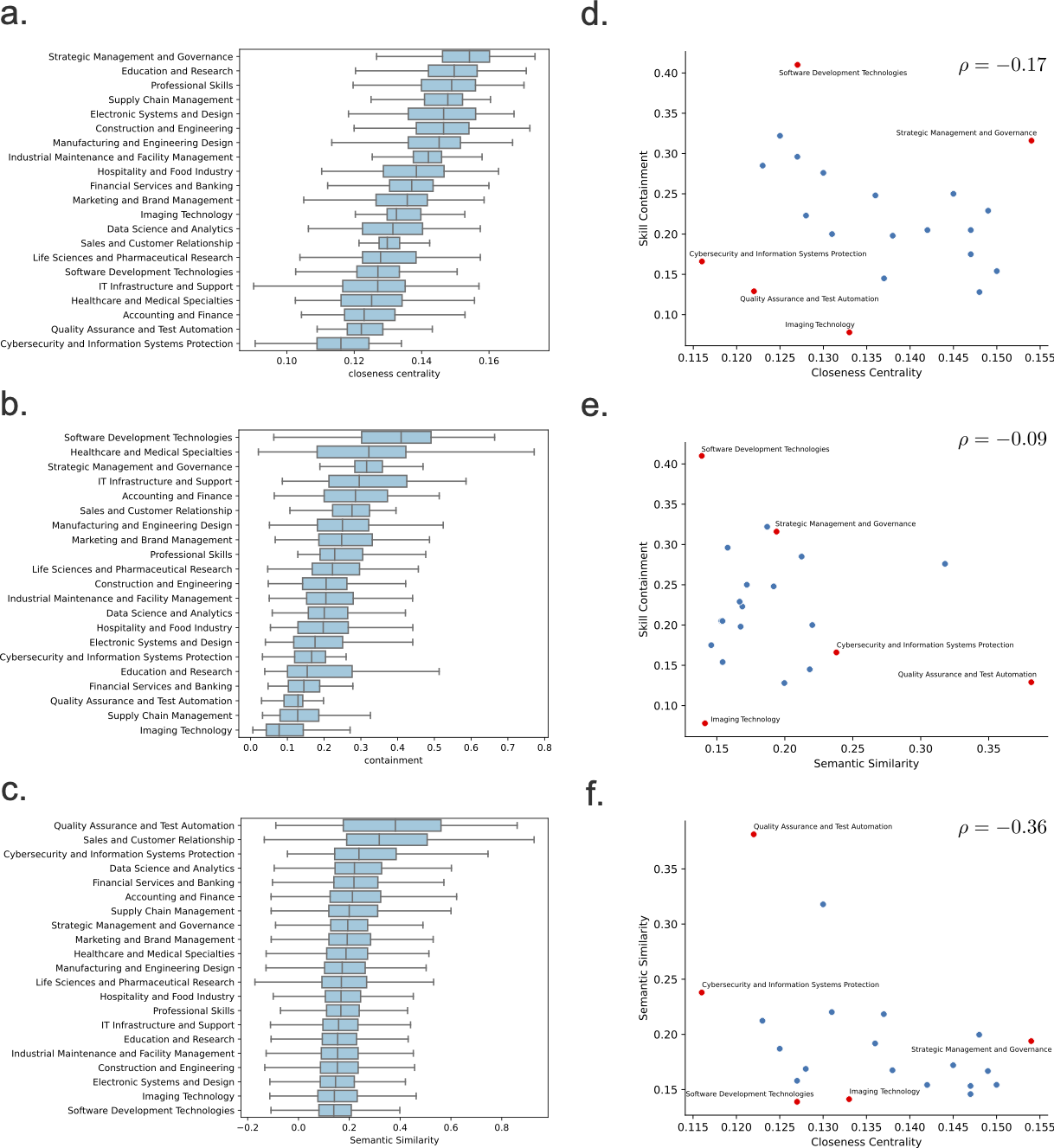}
    \caption{Boxplots for the distributions of  (a) closeness centrality (b) containment and (c) within cluster semantic similarity for each cluster. The scatter plots compare for each cluster: (d) median closeness centrality and containment, (e)  semantic similarity and containment, and (f) semantic similarity and closeness centrality.
    }
    \label{fig: centrality_containment}
\end{figure}

\begin{itemize}
\item \textit{{Average mentions:}}
Calculated as the number of mentions of skills from a skill cluster normalised by the number of adverts.
The average mentions range from common skills in the clusters `Strategic Management and Governance' (117 million mentions, 1.79 per advert) and `Professional Skills' (79 million mentions, 1.22 per advert) to the rarest skill clusters `Imaging Technology' (1.3 million mentions, 0.02 per advert), `Cybersecurity and Information Systems Protection' (2.5 million mentions, 0.04 per advert) and `Quality Assurance and Test Automation' (2.9 million mentions, 0.04 per advert). 

\item \textit{{Within-cluster semantic similarity:}}
Calculated as the median cosine similarity between the text embeddings of any two skills in the cluster computed from the same NLP model~\cite{reimers-2019-sentence-bert} used by Nesta for skill matching.
The diverse levels of semantic consistency of skills within each cluster captures differences in the homogeneity of co-occurrent skills across clusters . 
Figure~\ref{fig: centrality_containment}c shows that 
`Quality Assurance and Test Automation' (0.381) and `Sales and Customer Relationship' (0.318) exhibit high within-cluster semantic similarity, also confirmed by their word clouds in Figure~\ref{fig: MS21_fullinfo}. 
Conversely, `Imaging Technology' (0.141) and `Electronic Systems and Design' (0.146) have the lowest semantic similarity, signalling a grouping of more diverse skills.  As shown in Figure~\ref{fig: MS21_fullinfo}, `Imaging Technology' contains the quite different skills of `Landscaping' and `Medical Imaging', while `Electronic Systems and Design' spans seemingly diverse skills including `Electronics' and `Life Cycle Planning'. 
Therefore, the co-occurrence of these skills in job adverts is not aligned with the generic semantic understanding from language models, and could be linked to, e.g., technical content. Such discrepancies between skills co-occurrence and their semantic similarity may point towards emerging or innovative skills relationships, or to areas where more diverse skills are employed. 

\item \textit{{Skill cluster containment:}} 
Calculated for each node as the  weighted degree of a node within its cluster subgraph normalised by its weighted degree in the overall graph $\mathcal{G}$. 
A skill with high containment is more likely to co-occur with skills that belong to the same skill cluster. The median over the cluster quantifies the extent to which a skill cluster contains a consistent set of co-occurring skills.
As shown in Figure~\ref{fig: centrality_containment}b, high skill containment is observed for `Software Development’, `Healthcare and Medical Specialities’ and `Strategic Management and Governance’, all with median skill containment values above 0.3. 
At the other extreme, we have clusters `Imaging Technology’, `Supply Chain Management’, `Quality Assurance and Test Automation’ and `Financial Services and Banking’ all with median containment below 0.15.
Note that even for the most contained clusters, connections outside of the skill cluster are overall stronger than connections within, underscoring the interconnectivity of skills in UK job adverts, with no isolated skills silos. 
Further evidence is seen in Figure~\ref{fig:coarse_grained_coverage}, where we show the large extent of cross-coverage (i.e., low containment) of mentions across skill clusters. 

\begin{figure}[htb!]
    \centering
    \includegraphics[width = 0.6\textwidth, angle=0]{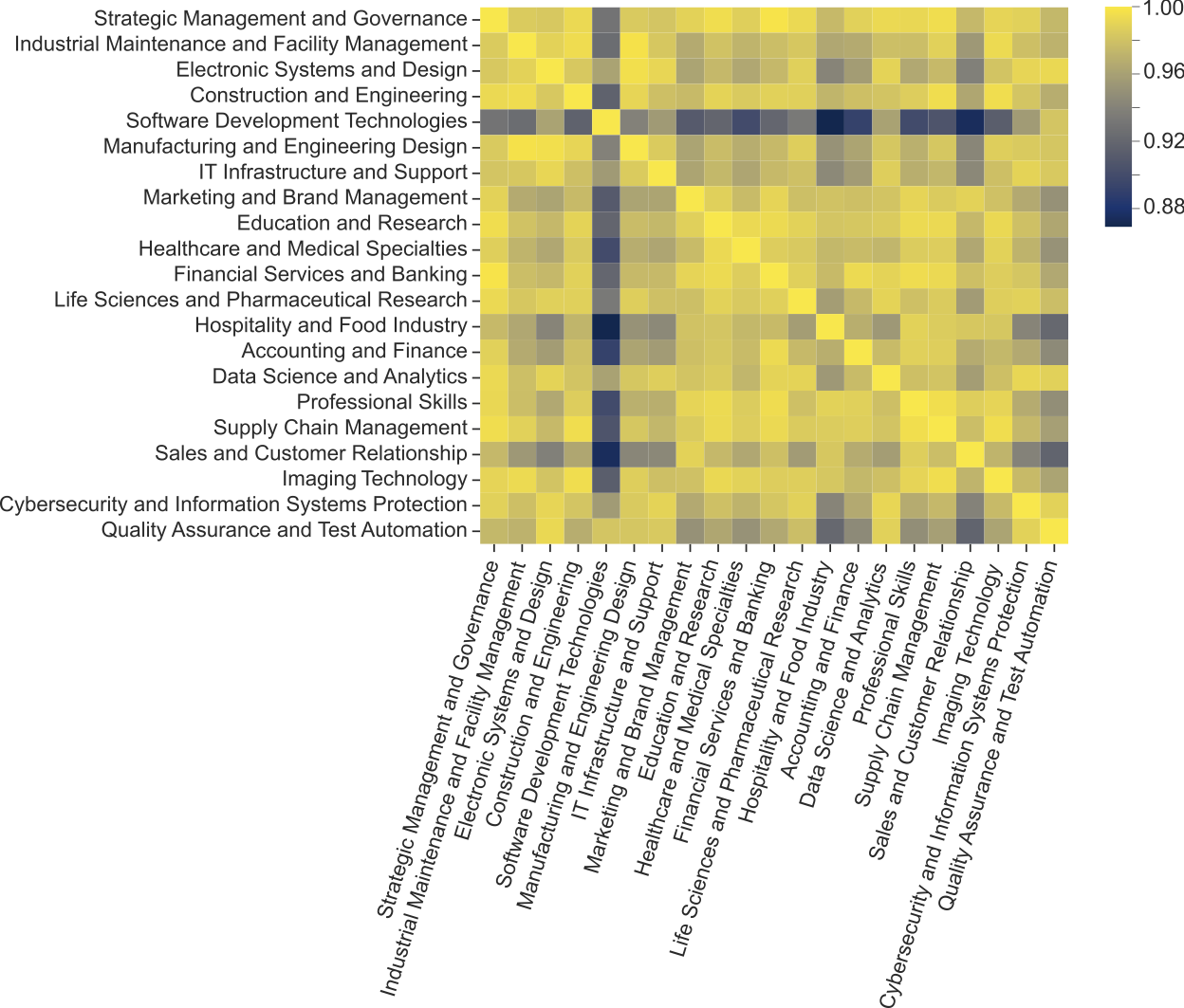}
    \caption{\textbf{Coverage between MS21 clusters.} 
  Coverage of the coarse-grained co-occurrence matrix $K$ obtained by: (i) computing $B=U^T K U$ where $U$ is the matrix where $U_{ij}$ that records the number of mentions of a skill $i$ in cluster $j$; (ii) normalising $B_{ij}$ by $\sqrt{B_{ii} B_{jj}}$. Coverage and containment have opposite meanings: the `Software Development Technologies' cluster has high self-containment (i.e., low values of its coverage), and is especially unlikely to co-occur with `Sales and Customer Relationship' or `Hospitality and Food Industry'.
    }
    \label{fig:coarse_grained_coverage}
\end{figure}

\item \textit{{Closeness Centrality:}}
Calculated as the average distance between each node and all others, along shortest paths on the graph $\mathcal{G}$~\cite{freeman2002centrality}. This was calculated for each skill using the NetworkX python package (version 3.2)~\cite{networkx} and we obtain the median for each cluster.
 A cluster with high closeness centrality is indicative of a group of skills that provide a core of common skills to access jobs from across the labour market, as they tend to co-occur in job adverts with a broad set of skills. 
Figure~\ref{fig: centrality_containment}a shows that closeness centrality appears to reflect a gradient from generic to specific skill clusters. For example, `Strategic Management and Governance’, `Education and Research’ and `Professional Skills’ have the highest closeness centrality, while `Cybersecurity and Information Systems Protection’, `Quality Assurance and Test Automation’ and `Accounting and Finance’ have the lowest closeness centrality, as they co-appear less frequently with other skills. 
Furthermore, individual skills with high closeness centrality within each cluster also correspond to more generic skills, a fact we used in our cluster labelling algorithm. 
\end{itemize}

\paragraph{Network properties and roles in the skills network.}
Figure~\ref{fig: centrality_containment} shows the lack of strong correlation between the semantic similarity, closeness centrality and containment of the skill clusters. The differences in these measures lead to distinct interpretations of the roles of the skill clusters in the skills network.  
For instance, clusters with high closeness centrality can be thought of as more `global’ in their relationships with other skills. On the other hand, skill clusters with high containment are comprised of ‘self-contained' skill sets. If we thought of `skill silos', we would thus expect relatively local and contained skill clusters, indicating a small set of separate skills in the network.
In our analysis, we find that `Software Development Technologies’ conforms with this expectation, as it is highly contained and also local. 
Other skill clusters have various characteristics. There are small and local clusters, such as `Cybersecurity and Information Systems Protection’, with low containment, i.e., skills that occur often alongside skills from other skill clusters, but do not have wide reach across the skills network. 
Conversely, we see that `Strategic Management and Governance’ is a large cluster that is `contained’, yet `global’. This indicates that there is a large number of skills in this core cluster that tend to co-occur with skills in the same cluster, and also permeate associations with skills across a wide range of other clusters. 

\begin{figure}[htb!]
    \centering
    \includegraphics[width=.95\textwidth]{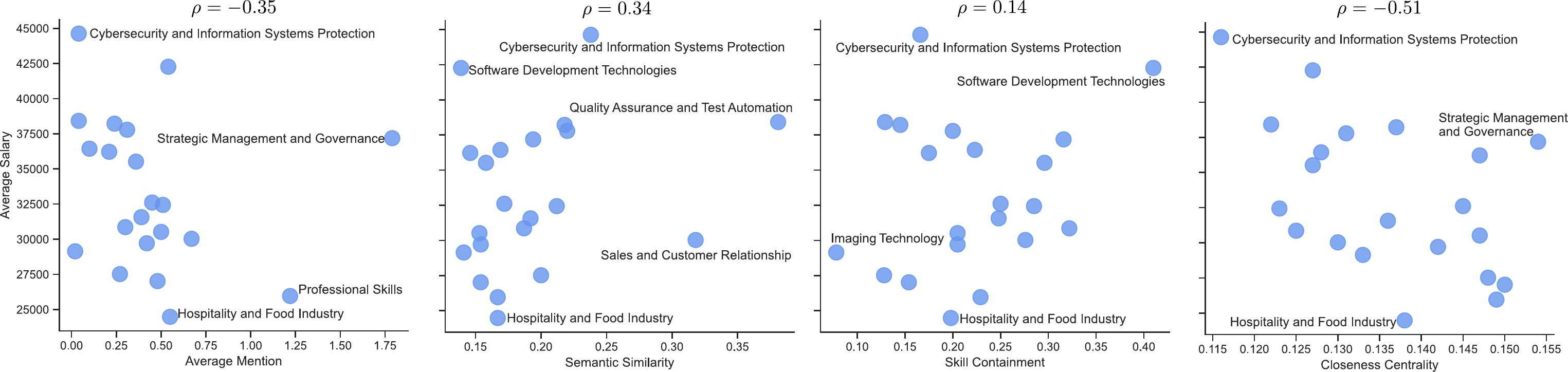}
    \caption{Scatter plots of average annual salary in £ \textit{vs.}\ (from left to right) average mentions, semantic similarity, skill containment and closeness centrality.
    }
    \label{fig:salary_corr}
\end{figure}

\paragraph{Average Salary and network properties.}
Table~\ref{tab: descriptive_table} also displays the mean predicted salary for each cluster, i.e., the average predicted salary of job adverts that mention a skill in that cluster.
`Cybersecurity and Information Systems Protection'  has the highest average salary (£44,630), followed by `Software Development Technologies' (£42,270). At the other extreme, adverts featuring skills from the `Hospitality and Food Industry' (£24,480) and `Professional Skills' (£25,960) clusters have the lowest average salary. 
Figure~\ref{fig:salary_corr} presents the pairwise relationships between salary and the network measures from Table~\ref{tab: descriptive_table}. Although the correlations between salary and network variables are weak, we observe a negative correlation between the average pay of a skill cluster and its median closeness centrality  (Spearman $\rho$ = -0.51) and the average mentions (Spearman $\rho$ =-0.34), and a positive correlation with the within-cluster semantic similarity (Spearman $\rho$ =0.32) . Together, these correlations hint at a salary premium afforded to skills that are more specialist and not commonly shared across the wider skills network.

\subsubsection*{The UK geography of skill clusters}

To analyse the geography of the MS21 skill clusters, we subsample the data set (every 11th advert ordered by date) keeping adverts with full information on skills, location and predicted salary. This results in 2.6 million job adverts uniformly spread across 2016, 2018, 2020 and 2022. Skills are assigned to adverts and adverts ascribed to NUTS2 regions, as described above. This allows us to compute the percentage of adverts in each NUTS2 region that are assigned to each skill cluster.

\begin{sidewaysfigure}
    \centering
    \includegraphics[width = 1.05\textwidth]{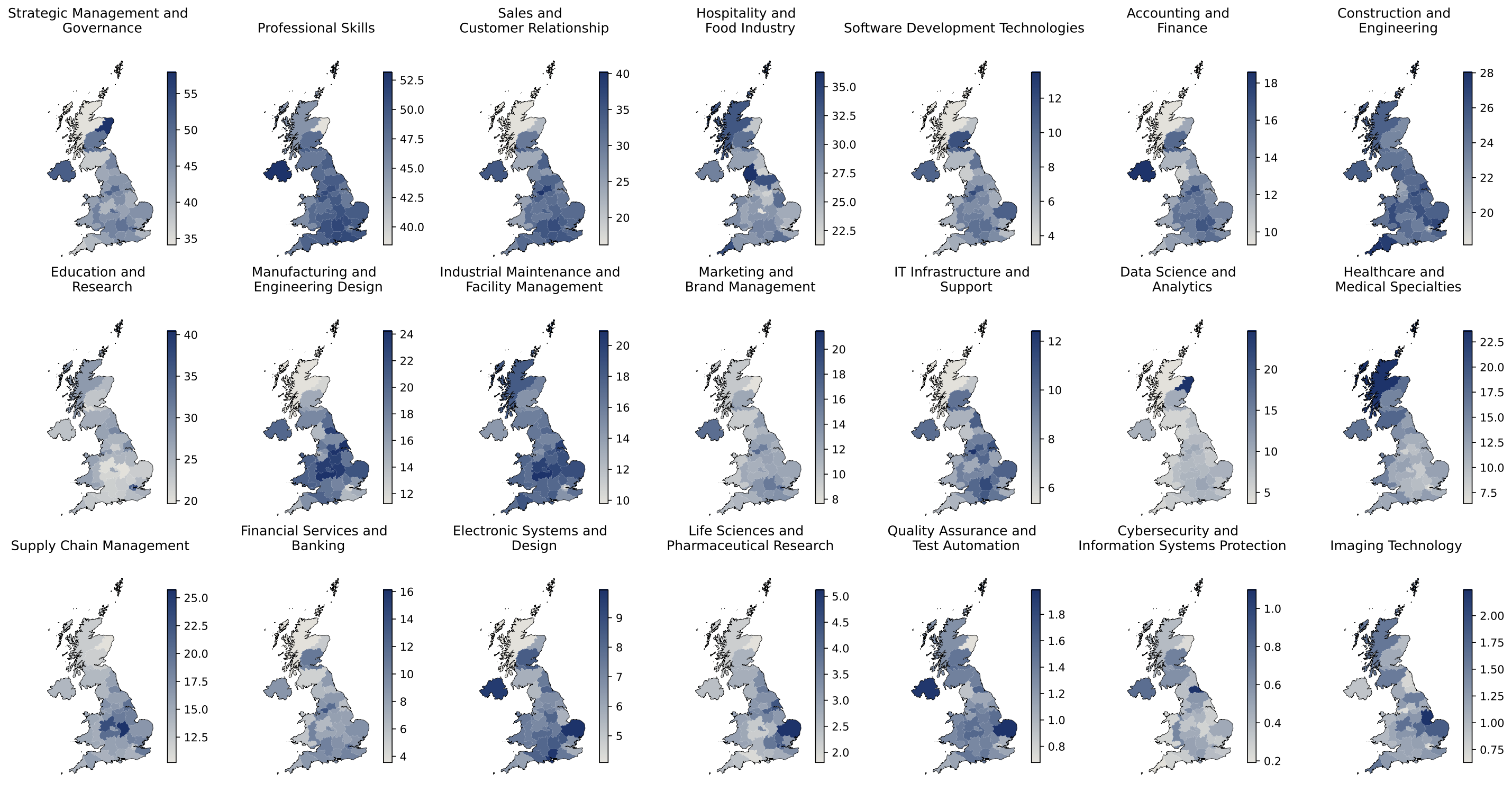}
     \caption{Maps showing the percentage of all adverts in each NUTS2 region featuring a skill from each of the MS21 skill clusters.
     } 
    \label{fig: 21_maps}
\end{sidewaysfigure}

\paragraph{Regional summary.}
Figure~\ref{fig: 21_maps} presents 21 maps (one for each MS21 skill cluster) showing the percentage of adverts that feature a skill from the cluster in each NUTS2 region. Clear geographic variation is evident, with `Strategic Management and Governance' being particularly common in North East Scotland, Northern Ireland, central London and in the conurbations of the West Midlands and Greater Manchester. Conversely, `Professional Skills' is particularly prominent in the counties in the commuter belt surrounding London, but also remains prominent in Northern Ireland. The important role of the hospitality industry in rural areas is shown by the prominence, percentage-wise, of `Hospitality and Food Industry' in the Highlands and Islands, Cumbria, North Yorkshire and Cornwall. Similarly, the relative prominence of `Healthcare and Medical Specialties' in the Highlands and Islands, Cumbria, Durham and Tees Valley and Somerset supports the role of public sector employment in these areas.

Northern Ireland stands out for its high proportion of adverts featuring skills from `Accounting and Finance', `Electronic Systems and Design', and `Quality Assurance and Test Automation', three largely technical, quantitative skills clusters.
Two of these clusters (`Electronic Systems and Design' and `Quality Assurance and Test Automation') also feature prominently in East Anglia, alongside `Life Sciences and Pharmaceutical Research', reflecting the world leading role that Cambridge, both through its university and nearby businesses, plays in the technology and life sciences industries. Other stand-out percentages significantly above  the average are seen for `Supply Chain Management' in Leicestershire, Rutland and Northamptonshire, `Data Science and Analytics' in North Eastern Scotland, and `Cybersecurity and Information Systems Protection' in Tees Valley and Durham. 

Central London features prominently in `Accounting and Finance', `Education and Research', `Marketing and
Brand Management' and `Financial Services and
Banking', reflecting its position as the financial centre of the UK, while also being home to many large universities and the headquarters of many national and international companies. 

\begin{figure}[htb!]
    \centering
    \includegraphics[width = .95\textwidth]{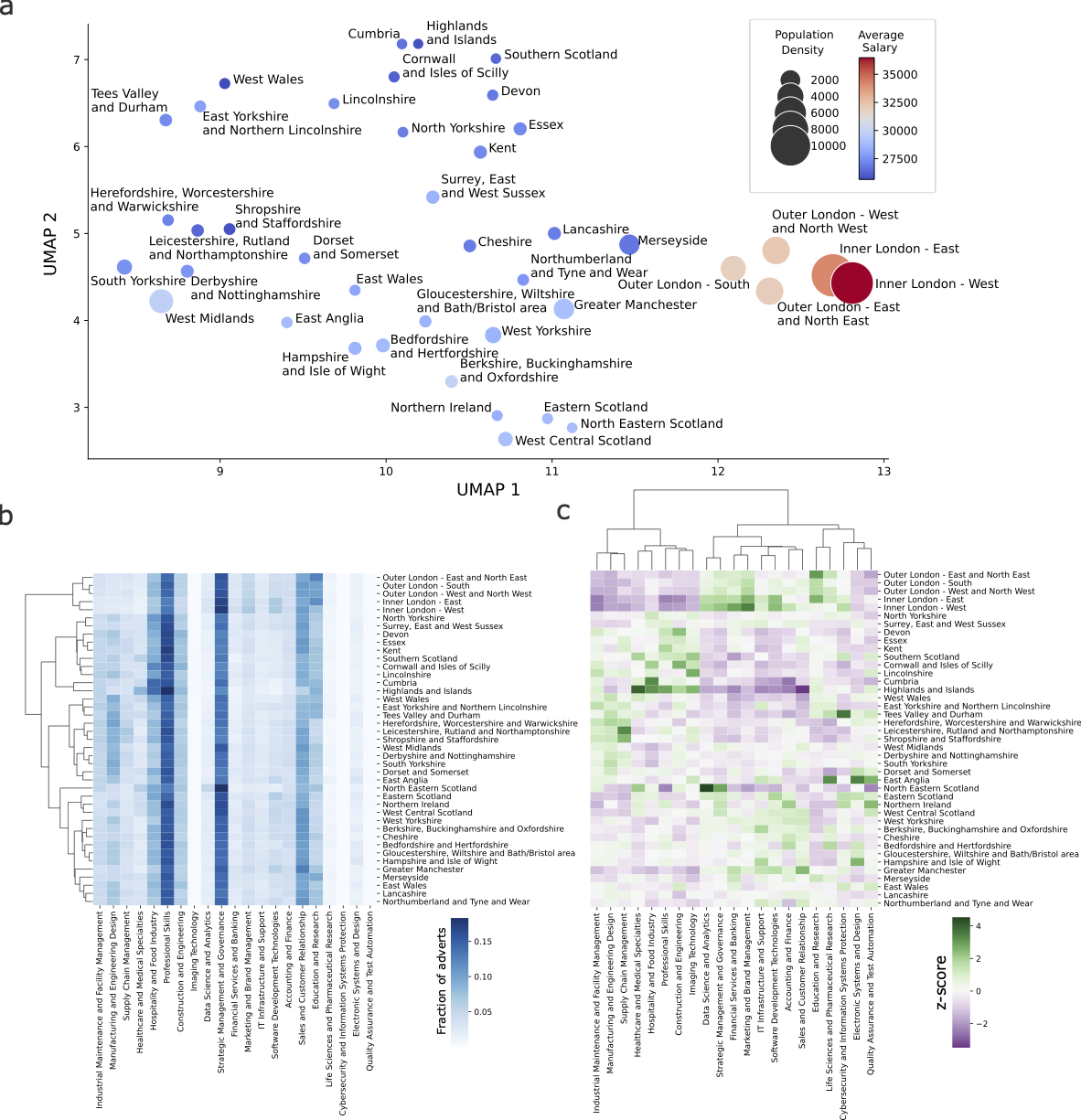}
     \caption{\textbf{Projection of the skill profiles of NUTS2 regions.} (a) UMAP projection of the set of 21-dimensional vectors of percentages in the MS21 skill clusters for all NUTS2 regions. NUTS2 regions with similar skill profiles lie close in this projection. The dots of the NUTS2 regions are then coloured \textit{a posteriori} according to average salary in the region. (b) Hierarchical clustering of the NUTS2 regions based on their MS21 percentages.  (c) Hierarchical clustering of the MS21 clusters based on z-scores across regions.
     } 
    \label{fig: UMAP_NUTS2}
\end{figure}

\paragraph{Dimensionality reduction.}
To further inform the geographical characterisation in Figure~\ref{fig: 21_maps}, we perform dimensionality reduction on the MS21 skill profiles of all NUTS2 regions (see Figure~\ref{fig: UMAP_NUTS2}b). Specifically, for each of the 41 NUTS2 regions, we consider the 21-dimensional vector with coordinates equal to the percentage of adverts featuring a skill from each of the MS21 skill clusters.  
We then project this set of vectors onto the plane using UMAP~\cite{mcinnes2018umap}, a nonlinear projection technique that preserves relative distances between high-dimensional vectors. As a result, NUTS2 regions with similar skill profiles are placed close to each other on the plane defined by the two components of the projection, UMAP1 and UMAP2 (Figure~\ref{fig: UMAP_NUTS2}a).

As expected, we find regional groupings that reflect shared geographic, sociodemographic, occupational and industrial similarities. Specifically, there is a distinct London grouping  (at large values of UMAP1), a cluster with the urban regions of Scotland with Northern Ireland (at low values of UMAP2) which lie close to affluent regions of the South of England, while the predominantly rural regions of England are grouped close together at large values of UMAP2 and the traditional industrial regions of England are  clustered at low values of UMAP1.  
%

Further details of the similarities and dissimilarities in regional skill profiles are presented in Figure~\ref{fig: UMAP_NUTS2}c, which shows, for each region, the z-score of the percentages of MS21 skill clusters, when compared across all regions, Hence this allows us to measure how much the percentage of adverts featuring a skill from a skill cluster for a given region deviates from the average observed across all regions. The hierarchical ordering of skill clusters highlights groups of skills that help understand the regional differences across the UK. 
In particular, most urban regions in England are grouped closely towards the center of the UMAP2 coordinate, and spread out along the UMAP1 coordinate. The variation along the UMAP1 coordinate captures a change in regional skill profiles that go from, on one side, higher than average percentages in skills related to manufacturing (`Manufacturing and Engineering Design', `Industrial Maintenance and Facility Management', `Supply Chain Management') in regions such as 
West Midlands,
South Yorkshire,
and
East Yorkshire and Northern Lincolnshire
to, on the other side, the London grouping, which is characterised by higher percentages of skills related to finance, large corporations, civil service, education, data science (e.g., 
`Data Science and Analytics'
`Financial Services and Banking'
`Marketing and Brand Management'
`Software Development Technologies',
`Accounting and Finance',
`Education and Research',
`Life Sciences and Pharmaceutical Research'). In between these two extremes lies the grouping of 
Greater Manchester,
Merseyside,
Lancashire, and
Northumberland and Tyne and Wear.

On the other hand, at large values of the UMAP2 coordinate  we find rural or tourist heavy areas, 
such as 
Cumbria,
Highlands and Islands,
Surrey, 
Kent,
Southern Scotland,
North Yorkshire,
and Cornwall and Isles of Scilly.
with dominance of skills such as
`Professional Skills'
`Hospitality and Food Industry' or
`Construction and Engineering'.
On the other extreme of the UMAP2 coordinate, we find a grouping of English regions including 
Berkshire, Buckinghamshire and Oxfordshire,
West Yorkshire
and Cheshire, 
which are characterised by higher percentages of skills in
`IT Infrastructure and Support',
`Accounting and Finance',
`Sales and Customer Relationship',
`Life Sciences and Pharmaceutical Research',
and
`Electronic Systems and Design'.
Further down the UMAP2 coordinate, we find the Scottish and Northern Ireland grouping 
containing 
Northern Ireland,
Eastern Scotland, and
West Central Scotland,
which also has higher percentages than average in  
`Software Development Technologies' and 
`Accounting and Finance', but also
higher percentages in skills related to 
`Strategic Management and Governance',
`Data Science and Analytics',
`Electronic Systems and Design'
`Quality Assurance and Test Automation', and 
`Cybersecurity and Information Systems Protection'.

The z-scores also highlight regions with particular skill percentages that deviate significantly from the average:
`Life Sciences and Pharmaceutical Research', `Electronic Systems and Design', and
`Quality Assurance and Test Automation' (East Anglia); 
`Strategic Management and Governance' and  `Data Science and Analytics' (North Eastern Scotland); 
`Supply Chain Management' (Leicestershire, Rutland and Northamptonshire);
and `Cybersecurity and Information Systems Protection' (Tees Valley and Durham).
These different mixes of skills are linked to sectors with varying levels of salaries, as seen in Figure~\ref{fig: UMAP_NUTS2}a, and will be the object of future research.









\subsubsection*{Temporal trends in the skill clusters}

Next we examine changes between the start and end of our temporal window, i.e., between 2016 and 2022.
To do so, we generate data sets for each year. We have a total of 15,861,000 adverts posted between 1st April 2016 and 31st December 2016, and 19,696,844 adverts posted between 1st January 2022 and 31st December 2022, equating to 57.89 and 53.96 adverts per day in 2016 and 2022, respectively. Overall, we find that 
the average number of skills per advert grew from 8.60 in 2016 to 10.73 in 2022, highlighting the increase in skills requirements within single job adverts.

\begin{figure}[htb!]
    \centering
    \includegraphics[width=0.95\textwidth]{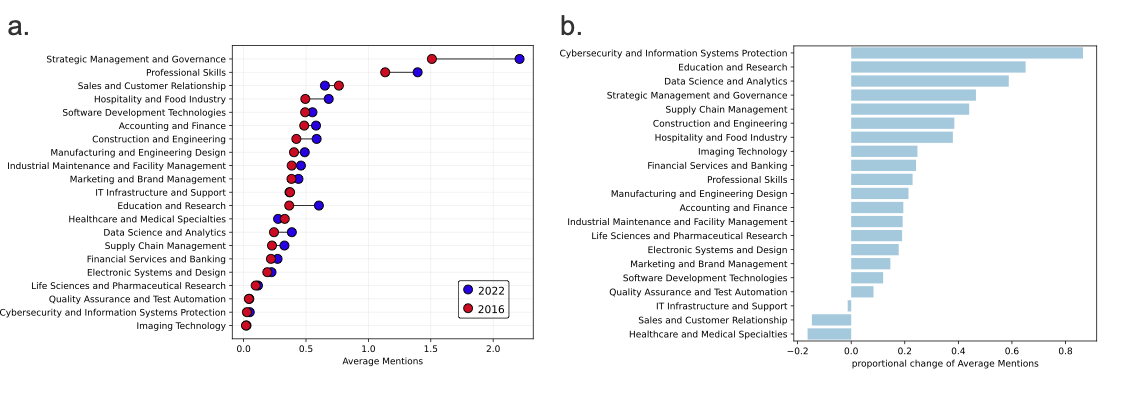}
    \caption{\textbf{Temporal changes of skill demand.} (a) Temporal changes in average mentions for all skill clusters from 2016 to 2022 and (b) the corresponding relative changes. 
    }
    \label{fig: temporal_volume_salary}
\end{figure}

Figure~\ref{fig: temporal_volume_salary} shows a general increase in the average mentions across the 21 skill clusters between 2016 and 2022. The largest increases in the mentions per advert were observed for `Strategic Management and Governance' (1.51 mentions per advert in 2016 to 2.21 mentions per advert in 2022), `Professional Skills' (1.13 to 1.39) and `Education and Research' (0.37 to 0.60), but large relative increases are observed for `Cybersecurity and Information Systems Protection' (86.51\%), `Education and Research' (65.10\%) and `Data Science and Analytics' (58.82\%). Decreases in frequency of mentions over this period were only found for `Sales and Customer Relationship' (-14.67\%) and `Healthcare and Medical Specialties' (-16.28\%). 

\begin{figure}[htb!]
    \centering
    \includegraphics[width=0.95\textwidth]{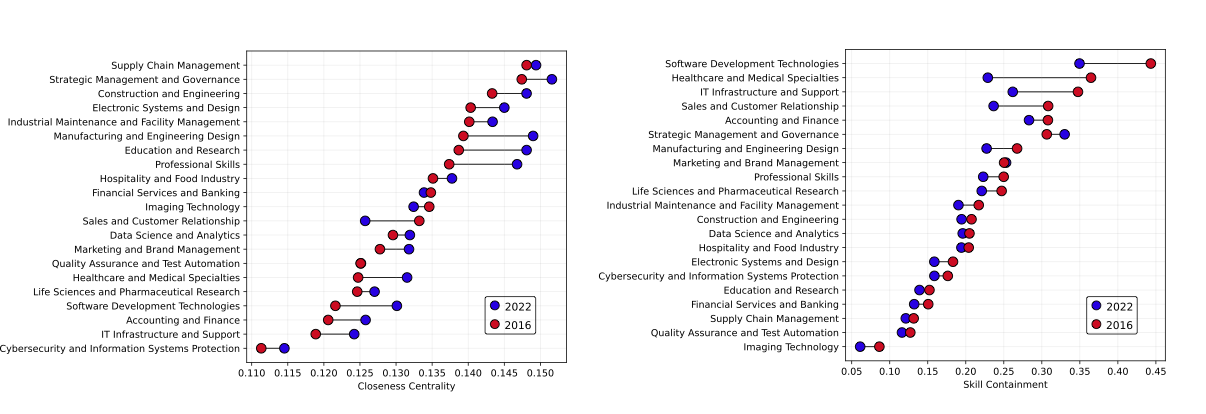}
    \caption{\textbf{Temporal changes of skill centrality and containment.} (a) Temporal change in closeness centrality and (b) skill containment from 2016 to 2022 for all skill clusters. 
    }
    \label{fig: temporal_closeness_containment}
\end{figure}

We also compare the closeness centrality and containment of skill clusters between 2016 and 2022. To do so, we  construct two different skills networks (i.e., two weighted sparsified graphs $\mathcal{G}_{2016}$ and $\mathcal{G}_{2022}$, as described in Methods) , 
and we compute network summary properties as described above.
Figure~\ref{fig: temporal_closeness_containment}a shows that closeness centrality increased from 2016 to 2022 in 18 of the 21 skill clusters, with the exception of `Sales and Customer Relationship', `Imaging Technology' and `Financial Services and Banking'. The largest increases in centrality were found in `Manufacturing and Engineering Design', `Professional Skills', and `Education and Research'. 
Conversely, Figure~\ref{fig: temporal_closeness_containment}b shows that skill containment fell between 2016 and 2022 in 19 of 21 skill clusters, confirming the emergence of stronger cross-cluster relationships over time and a broadening of skills requirements (Figure~\ref{fig: temporal_closeness_containment}). 
Particularly large decreases of skill containment are noted for `Software Development Technologies', `Healthcare and Medical Specialties' and `IT Infrastructure and Support'. This may indicate that jobs requiring these skills are transitioning from specialist roles in 2016 to less specialist roles or roles spanning multiple skill groups. These observations may also indicate the growing requirement for skills featured in these clusters as supplementary skills in roles whose core skills are in other skill clusters; for example a requirement for knowledge of a computer coding language for a sales role.

Overall, these observations indicates that between 2016 and 2022 the skills requirements of job adverts have become more overlapping, with job adverts more frequently requiring skills spanning skill clusters. These findings align with recent work examining the rate of turnover of skills in the UK labour market, in which an increasing breadth of skills required for many roles was observed ~\cite{rohenkohl_old_2024}.

\subsubsection*{Contrasting data-driven skill clusters with expert-based skill categories}

It is interesting to contrast our data-driven skill clusters, which have been directly derived, agnostically, from their co-occurrence in job adverts,  to expert-based classifications of skills into categories, some of which include the Lightcast Open Skills Taxonomy and the OECD Skills for Jobs database~\cite{lightcast_skills, oecd_skills}. 
Given that individual Adzuna skills have been already matched to LC skills (see Methods), we examine directly the correspondence between our data-driven skill clusters (MS21) and the expert-based LC skills categories (32 categories).

\begin{figure}[htb!]
    \centering
    \includegraphics[width=.67\textwidth]{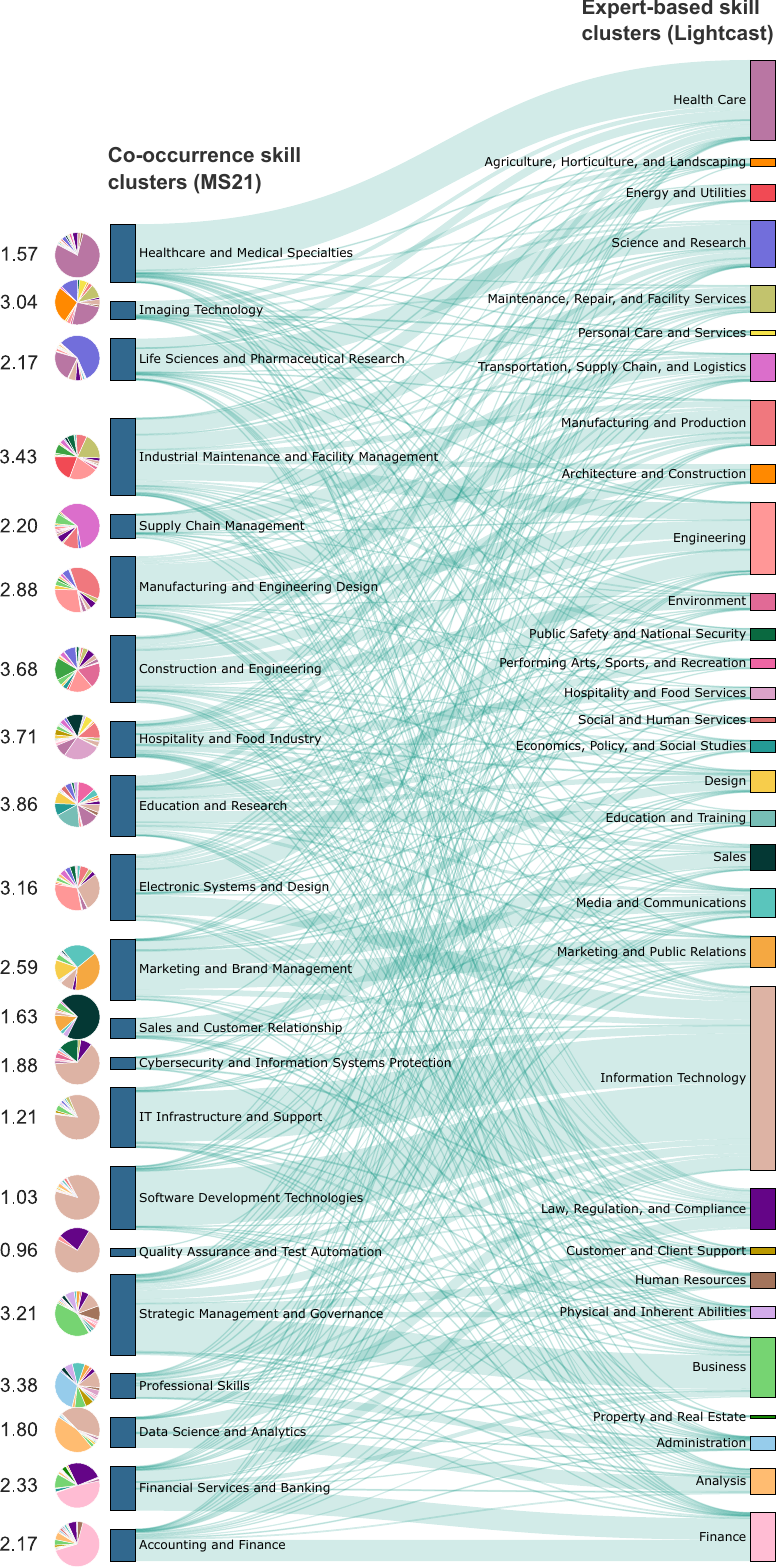}
    \caption{\textbf{Sankey diagram between co-occurrence skill clusters (MS21) and expert-based skill categories (Lightcast)}. There is some agreement between the data-driven clusters and the LC categories in skill areas where thematic content and co-occurrence match. For each cluster in MS21, we plot a pie chart to visualise the proportions of Lightcast categories, and the corresponding `thematic entropy' for the skill cluster, which indicates how thematically mixed the cluster is.} 
    \label{fig: SANKEY_21lightcast}
\end{figure}

Figure~\ref{fig: SANKEY_21lightcast} shows broad agreement between MS21 and LC but not uniformly across all groupings. This difference is expected and indicative that the sets of skills required by employers in an advert often span diverse thematic categories. 
In particular, the LC category 'Information Technology' is too broad to capture the variety of relationships in the skills co-occurrence network. Hence this group of skills is spread across several MS21 skill clusters, most notably `Software Development Technologies’, `IT Infrastructure and Support’, `Electronic Systems and Design’, `Cybersecurity and Information Systems Protection’ and ‘Data Science and Analytics’. Notably, these skill clusters correspond partly to a finer level of the LC taxonomy (sub-categories), and indicates the importance of using intrinsic scales in the process of clustering to capture the natural associations in the data.

Some of the MS21 clusters span several Lightcast categories, 
as shown by the large values of the thematic entropy values and pie charts in Figure~\ref{fig: SANKEY_21lightcast}: `Education and Research' (entropy = 3.86), `Hospitality and Food Industry' (3.71) and `Construction and Engineering' (3.68) map to several LC categories. Conversely, other skill clusters, such as `Quality Assurance and Test Automation' (0.96), `Software Development Technologies' (1.03) and `IT Infrastructure and Support' (1.21) all map closely to one Lightcast category (`Information Technology'), and thus have low entropy values. 

As expected, MS skill clusters have lower semantic similarity than LC skills categories (LC: 0.234 larger than MS21: 0.172).
This is unsurprising, and follows directly from the fact that our MS21 skill clusters emerge from co-occurrence in adverts, thus reflecting the need for dissimilar skills in certain jobs, whereas LC categories follow from expert knowledge, hence partly based on thematic and semantic content.


\newpage
\section*{Discussion}

Using data from 65 million job adverts in the UK between 2016 and 2022, we use a network construction and graph-based multiscale clustering to find data-driven skill clusters based on their co-occurrence patterns, as demanded by employers. 
Our analysis has focused on a configuration of 21 skill clusters (MS21), identified as optimal based on data-driven criteria, as providing enough granularity and interpretability. 


To analyse the relationship between skills in the co-occurrence network, we use three metrics (closeness centrality, skill containment, semantic similarity), which allow us to quantify the level of participation of skill clusters within, and outside, their own group, as well as evaluating the level of thematic consistency of the skill clusters. 
We find that the skill clusters in MS21 have different roles within the  network. Some clusters have strong relationships with a small number of other clusters, while others tend to occur frequently with a broad range of other skills. `Cybersecurity and Information Systems Protection' is notable for often occurring alongside skills from other clusters, but has less reach across the skills network as a whole, suggesting its role as a supporting skill across sectors. Conversely, `Strategic Management and Governance' is more likely to co-occur with skills from within its cluster, but has wide reach across the wider skills network, suggesting its role as a necessary skill for jobs across a broad range of disciplines. We find a moderate negative correlation between the closeness centrality of a skill cluster and the average pay of adverts in which it features, suggesting a pay premium for less common, more specialised skills. 

We find notable differences in the geographic distribution of skill clusters across England. The two most common clusters of `Strategic Management and Governance' and `Professional Skills' have quite different spatial distributions, while less common clusters are found in specific regions where their skills are particularly in demand. This largely reflects variation in the industrial and occupational composition of these regions and will be studied in further work. 

Between 2016 and 2022, we find evidence that a wider diversity of skills is being required in job adverts: on average, adverts are now spanning more skill clusters. Overall the closeness centrality of skills increases, while the within-cluster skill containment decreases. Notable decreases in the containment and increases in the closeness centrality of `Software Development Technologies' suggests  previously contained technical skills being more widely required across the job market.

When we compare our skill clusters to the thematic categories in the Lightcast Open Skills taxonomy, we find partial agreement, suggesting our method may offer a different way to group skills based on observed usage rather than prescribed expert categories based on competencies and sectors. 
This research opens up several areas of future research.
Firstly, the hard partitioning of skills into mutually exclusive, collectively exhaustive clusters might not be the most adequate way to reflect the highly interconnected nature of skills co-occurrence patterns. This could be ameliorated with the use of soft, local partitioning methods~\cite{yu2020severability} that reflect more faithfully the overlaps of skills.
A further area of research is the use of additional network measures that capture the differences between core and periphery in the skills network~\cite{mucha2017coreperiphery}.
Another direction of work would involve the evaluation of the diversity and synergy in the skills in a job advert, as a means to characterise skills and occupations that enable transitions and evolution across jobs, as well as connecting further the geographical aspects of the analysis using further socio-economic data.   


The relationships between skills in the UK labour market are complex, and the demand for skills differs significantly across the country. Groups of skills are commonly required alongside one another in ways that are not expected based on the categorisation of skills by experts.
Over time, the dynamics of the UK skills network suggests a broadening of the skills required of workers in the UK,  with diverse skills being required together more often. 

\section*{Methods}
\subsection*{The UK job postings data set}

The data is provided by Adzuna Intelligence, an online job search engine that collates and organises information from various sources (e.g., employers' websites, recruitment software providers, traditional job boards), and generates a weekly snapshot that captures over 90\% of all jobs being advertised in the UK~\cite{bassier_vacancy_2023,adzuna_link}.
The original data set contained 197 million job adverts published by 606,450 different organisations and collected via weekly snapshots during 2016 (April-December, 9 months) and  2018, 2020 and 2022 (complete years), for a total of 45 months. 
Each job advert contains the free text of the original job description, and structured information scraped from the text, e.g., the date the advert was made available, and the name and location of the organisation posting the advert, 
among others. 
For this work, we extract from each job advert its unique identifier, date of first posting, and location associated with the advert, as well as two fields provided by Adzuna Intelligence's proprietary algorithms: the skills associated with each job, and the predicted salary, as discussed below.

\paragraph{Matching Adzuna skills to the Lightcast taxonomy:} 
To identify the skills present in each advert, Adzuna match specific keywords in the text of an advert to a dictionary of 6265 
pre-defined skills. To aid comparisons to other work, we map the skills extracted by Adzuna Intelligence onto the Lightcast (LC) Open Skills taxonomy~\cite{lightcast_skills}. The LC taxonomy is a hierarchy of skills, which has been used previously to study the changing skills requirements of science and technology jobs and the relationship between the skills demands of firms and their performance~\cite{deming_earnings_2020, deming_skill_2018}. 
The mapping to the LC taxonomy proceeds in two stages.  First, we use Nesta’s Skills Extractor v1.0.2~\cite{nesta} to match each Adzuna skill to the semantically closest LC taxonomy term, as measured by the cosine similarity between word embeddings computed using huggingface’s sentence-transformers/all-MiniLM-L6-v2 pre-trained model~\cite{reimers-2019-sentence-bert}. 
After this first step, 6265 unique Adzuna skills are matched to 4067 Lightcast skills.
Second, we apply manual curation and validation by expert researchers in our team to check terms, correct mismatches and enhance the quality of the matched pairs, including dropping generic or ambiguous skills against the taxonomy, e.g., terms that appear in recruiter
disclaimers (‘Luxury’, ‘Answering’, ‘Discrimination’,‘Dynamics’), or describers of the conditions and benefits of a job (‘Dental Insurance’, ‘Working abroad’, ‘Temporary Placement’). In total, a further 519 Adzuna skills are dropped. After the curation step, 5746 Adzuna skills are assigned to 3906 Lightcast skills. 
%
%

\paragraph{Predicted salary:}
The predicted salary of each advert is calculated by Adzuna Intelligence using a proprietary algorithm---a neural network trained 
to predict ground truth salaries, provided by the employers, from the job description, location of the role, contract type, and employing company. Note that the date of posting is not part of this model, hence we do not consider temporal changes in salary in this paper.  
%
To validate the predicted salaries against external data, the median predicted salary of each 2-digit SOC occupational code from the Adzuna dataset was compared to the corresponding  median salary from the Annual Survey of Hours and Earnings (ASHE) of the UK ONS for 2016 and 2022. ASHE data were  adjusted to 2016 prices according to the Consumer Prices Index ~\cite{cpi} to account for inflation.  As shown in Figure \ref{fig:ashe_comparison}, there is close agreement between Adzuna predicted salaries and ASHE salaries in both 2016 (Spearman's $\rho$ = 0.87) and 2022 (Spearman's $\rho$ = 0.90). In the lowest paid occupations, the Adzuna predicted salary was consistently higher than expected from official statistics, suggesting more highly paid positions within these occupations are more likely to be included in the Adzuna dataset. 
%

\paragraph{Deduplication of adverts:}
Given that job postings are compiled from several sources, and that postings can stay online for over a week, this can result in duplicated adverts.
Therefore we filter the adverts such that
each job advert (unique id) is included only once, using the first instance when the advert appears.
After this deduplication step, our data set contains 65 million unique adverts, of which 99.6\% contain at least one skill. 

\paragraph{Mapping of location data:}
We conduct geographical analyses at the level of NUTS2 (also known as ITL2) regions, 
corresponding to 41 non-overlapping regions in the UK 
 plus the country of Northern Ireland, with populations between 800,000 and 3 million residents. 
The locations are scraped directly from the free text and structured data in the job advert, and can correspond to locations that either fit within one NUTS2 region 
or that span more than one NUTS2 region (e.g. `London' spans five NUTS2 regions).
Adverts with raw locations contained entirely within a NUTS2 region are assigned to that region. 
%
For adverts with raw locations spanning more than one NUTS2 region, we allocate the advert at random to one of the spanned regions with probability given by the proportions of adverts in each region.

\subsection*{Constructing the skills co-occurrence network}\label{sec: network_representation}


\paragraph*{The co-occurrence matrix:}

Using the curated and deduplicated data set containing 65 million job adverts collected weekly during
2016, 2018, 2020 and 2022 we compile an $N \times N$ matrix $K$ of co-occurrence counts, where $N=3096$ is the number of unique skills  and $K_{ij}=m$ if a skill $i$ has co-occurred with skill $j$ in $m$ adverts. 

\paragraph*{Graph construction:}

\begin{figure}[htb!]
    \centering
    \includegraphics[width=0.75\textwidth]{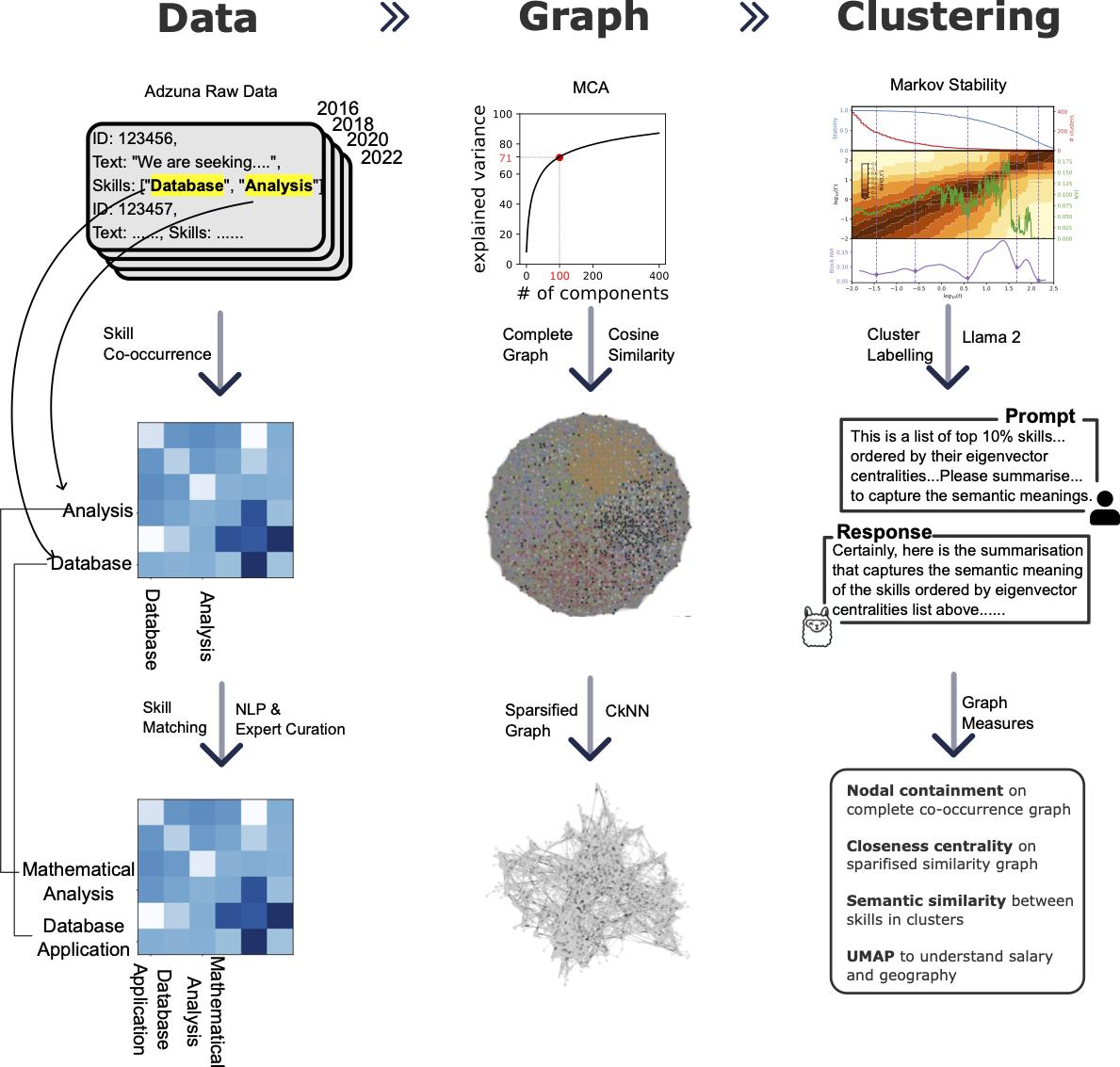}
    \caption{\textbf{From job postings to the clustering of a network of co-occurrent skills: } (a) the data preparation including the extraction of skill co-occurrence from metadata, skill matching to Lightcast taxonomy and dimensionality reduction using MCA; (b) the graph-based clustering including the sparsification of the complete cosine similarity graph and the multiscale clustering with Markov Stability; and (c) the descriptive analysis on the optimal clustering with 21 partitions using LLM, nodal containment and closeness centrality.   
    } 
    \label{fig:flowchart}   
\end{figure}

We then follow a graph construction protocol to obtain  a skills co-occurrence network, where the nodes of the network are skills and the edges of the graph connect skills with similar patterns of co-occurrence. To do this, we proceed as follows.

First, as is customary with sparse and noisy count matrices, we apply dimensionality reduction to project $K$ onto a lower dimensional space. Here, we use Multiple Correspondence Analysis (MCA)~\cite{le2010mca},
a multivariate extension of Correspondence Analysis, which is similar to Principal Component Analysis but appropriate for discrete variables. 
%
We apply the MCA dimensionality reduction to the skill co-occurrence matrix $K$ and obtain the first 100 MCA components, which explain 70\% of the variance of the original  data. The resulting MCA embedding is a set of 3096 embedding vectors (one for each skill, each of dimension 100) denoted $\{\mathbf{s}_i\}_{i=1}^{3096}, \, \mathbf{s}_i \in \mathbb{R}^{100}$ . Each vector provides, for each skill, a filtered, robust description of the leading co-occurrence patterns in the data. 

To measure the similarity between skills, we then compute $S$, the matrix of cosine similarities between the MCA embedding vectors of skills, where $S_{ij}= (\mathbf{s}_i /||\mathbf{s}_i||) \cdot (\mathbf{s}_j/||\mathbf{s}_j||)$. Although this (full) similarity matrix could be used directly for clustering, it has been shown that a graph formulation can be advantageous to enhance clustering for such high-dimensional, noisy data~\cite{liu2020cknn}.
To do this, note that the similarity matrix $S$ can be thought of as the adjacency matrix of a fully connected weighted graph, $\mathcal{G}_S$. However, such a graph contains many edges with small weights reflecting weak similarities---in high-dimensional, noisy data sets even the least similar nodes can present a substantial degree of similarity. Such weak similarities are in most cases redundant, as they can be explained through stronger pairwise similarities present in the graph~\cite{altuncu2019free,beguerisseRoleBasedRMST2013}. 

To reveal the intrinsic structure of the data, we sparsify the fully connected graph $\mathcal{G}_S$ by eliminating redundant edges through a geometric graph construction. We start by transforming similarities into distances  
$\widetilde{d}_{ij} = 1-S_{ij}$ and max-normalise to get
$d_{ij}=\widetilde{d}_{ij}/d_\text{max}$ where 
$d_\text{max}=\text{max}_{ij}(d_{ij})$ 
to ensure that the entries are bounded between 0 and 1~\cite{altuncu2019free}.
We then generate a sparsified geometric graph using Continuous k-nearest neighbours (CkNN)~\cite{Sauer_CkNN}, where two nodes $i$ and $j$ are connected if $d_{ij}<\delta \sqrt{d_i^k d_j^k}$, where $d_i^k$ is the distance between node $i$ and its $k$-th nearest neighbour and $\delta$ is a parameter. This construction has been shown to preserve consistent neighbourhoods (i.e., the similarities) in the data, yet correcting for the local density and eliminating redundant weak similarities~\cite{liu2020cknn}.
Here we use $\delta=1, k=15$ to produce a sparse geometric graph $\mathcal{G}_{CkNN}$, which maintains 22421 edges out of the 4039753 edges in $\mathcal{G}_S$.
The edges present in $\mathcal{G}_{CkNN}$ are weighted with the similarities $S_{ij}$ (or distances $d_{ij}$) to produce our final sparsified weighted undirected similarity (or distance) graph $\mathcal{G}$ with adjacency matrix $A$. 
A sketch of the process of graph construction is presented in Figure~\ref{fig:flowchart}.

\subsection*{Multiscale graph-based clustering of skills}

We use Markov Stability (MS)~\cite{
delvenne2010stability,schaub2012markov, delvenne2013stability} as implemented in the Python package PyGenStability~\cite{arnaudon2023pygenstability} to obtain robust communities in the skills network $\mathcal{G}$ at different levels of resolution.
MS naturally scans across levels of resolution to identify communities within which random walkers remain contained over extended periods.  This process uncovers a sequence of robust, optimised partitions of increasing coarseness.
MS was run over Markov scales that render between 4 and 400 clusters. 
We computed partitions at 720 scales, running 800 optimisation evaluations of the Leiden algorithm~\cite{traag2019leiden} at each scale. We selected 400 optimisations to compute the Normalised Variation of Information (NVI) at each scale.
For further details about Markov Stability see Appendix~\ref{app:MS}.

\paragraph{Assignment of adverts to clusters:}
A job advert may have several associated skills that collectively may span more than one skill cluster; hence there is not a one-to-one relationship between each advert and a skill cluster. Here we assign an advert to a cluster if it has at least one skill in that cluster, as in Ref.~\cite{stephany_what_2024}. 
Hence a single advert can be assigned to multiple clusters if it contains one or more skills from these clusters. This assignment of adverts to skill clusters is used below to calculate the average salary 
and the geographic distribution of each skill cluster.

\paragraph{Automated summary labels for skill clusters.}%
The \textit{a posteriori} interpretation of clusters obtained through unsupervised methods is a fundamental challenge, which is typically tackled using expert knowledge, a process that can be expensive, time-consuming and highly subjective.  
To aid the interpretability of our data-driven skill clusters, we implement an automated approach that exploits both the semantic representations of the skills (nodes) and the network structure in each skill cluster (subgraph). Specifically, we select the top 10\% nodes (or the top 20 nodes, whichever is larger) in each cluster based on the node eigenvector centrality computed from its cluster subgraph. This subset of nodes (skills) capture the core of the skill cluster.  
We then use a state-of-the-art large language model, Llama 2 70B~\cite{touvron2023llama} to summarise in a short phrase the semantic meaning of the selected subset of skills for each cluster using the following prompt: \textit{This is a list of the most representative skills extracted from a skill cluster and they are ordered by their eigenvector centralities in descending order. Please summarise the following list in one word or phrase such that it captures the semantic meaning of each skill. The list is: [`skill1', `skill2', ...].} 
The resulting summary phrases were then checked manually for consistency by experts in our team, also using word clouds.  
These labels are used throughout this paper to describe the skill clusters (see, e.g.,  the Sankey diagram in Fig.~\ref{fig: sankey_allconfig}).


\section*{Acknowledgements}

The authors thank Christopher Pissarides, Abby Gilbert, Thomas Beaney, Dominik J. Schindler, Meghdad Saeedian and Robert L. Peach for valuable discussions. We are also grateful to colleagues at Adzuna, particularly Scott Sweden and James Neave, for supplying the data used in this report. This work was done under the Pissarides Review into the Future of Work and Wellbeing, led by Professor Sir Christopher Pissarides (Institute for the Future of Work and London School of Economics). The Pissarides Review into the Future of Work and Wellbeing is a collaboration between the Institute for the Future of Work (IFOW), Imperial College London and Warwick Business School. Zhaolu Liu, Bertha Rohenkohl and Mauricio Barahona gratefully acknowledge support from the Nuffield Foundation. Mauricio Barahona also acknowledges support by the EPSRC under grant EP/N014529/1 funding the EPSRC Centre for Mathematics of Precision Healthcare at Imperial. Jonathan Clarke acknowledges support from the Wellcome Trust (215938/Z/19/Z). The views expressed herein are those of the authors and do not necessarily reflect the views of the Nuffield Foundation.

\printbibliography

\newpage
\appendix

\section*{Appendices}

\section{Comparison of predicted salaries and official wage statistics}

\begin{figure}[htb!]
    \centering
    \includegraphics[width=0.5\textwidth]{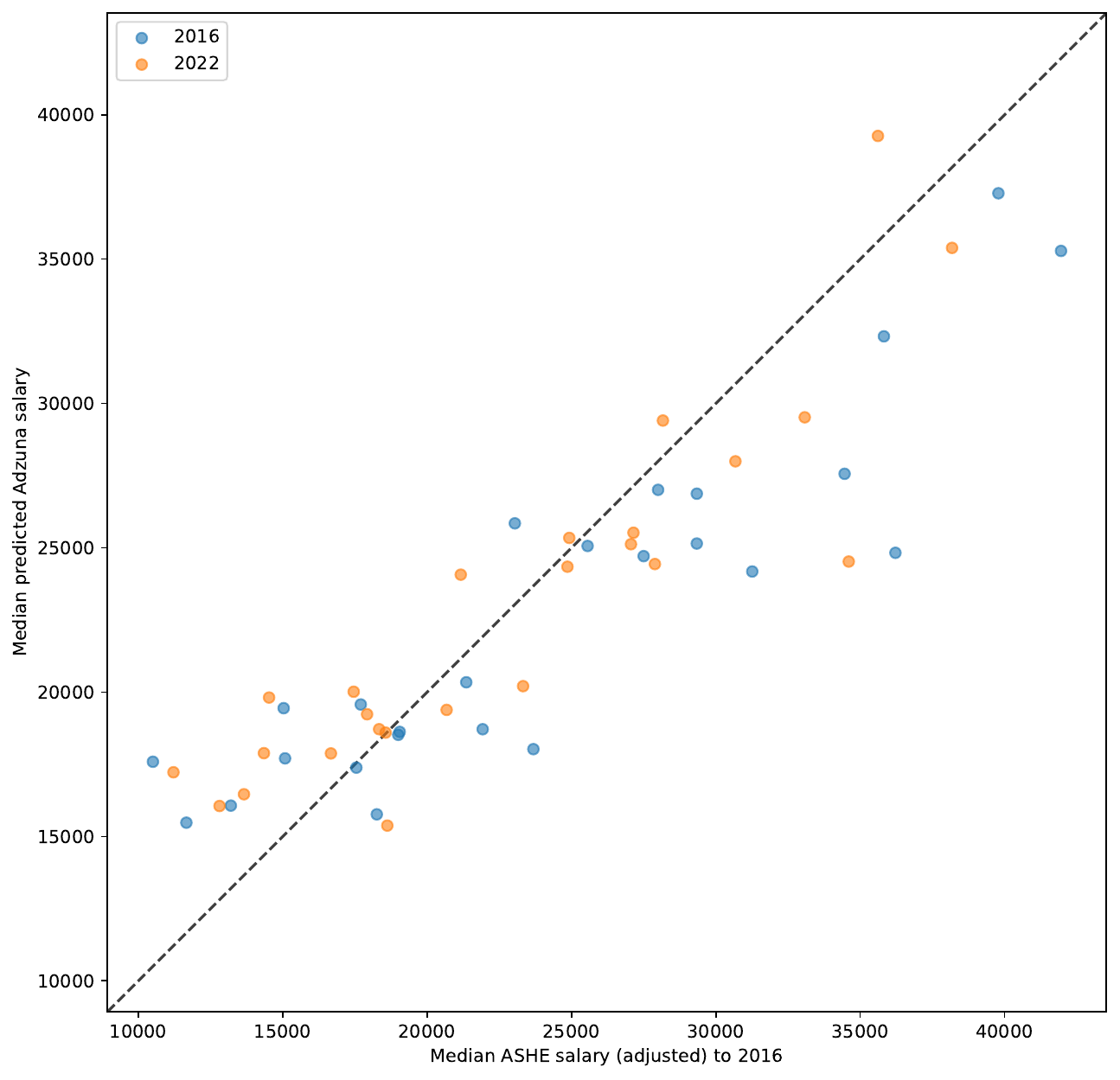}
    \caption{Comparison between median salary adjusted for inflation to 2016 prices from the Annual Survey of Hours and Earning (ASHE) and unadjusted Adzuna predicted salary for each two-digit SOC occupation code.
    } 
    \label{fig:ashe_comparison}   
\end{figure}



\newpage
\section{Multiscale community detection with Markov Stability}
\label{app:MS}

The skills graph $\mathcal{G}$ is analysed using Markov Stability (MS), a multiscale community detection framework that uses graph diffusion to detect communities in the network at multiple levels of resolution. For a fuller explanation of the ideas underpinning the method, see Refs.~\cite{delvenne2010stability, lambiotte2008laplacian, delvenne2013stability,liu2020cknn}.

Let $A$ be the $N \times N$ adjacency matrix and $D$ be the diagonal degree matrix of a graph $\mathcal{G}$. The transition probability matrix $M$ of a discrete-time random walk on $\mathcal{G}$ is:
\begin{equation}\label{S_eq:transition_prob}
    M = D^{+} A,
\end{equation}
where $D^{+}$ denotes the pseudo-inverse of $D$.
The matrix $M$ defines a discrete-time Markov chain on the nodes of $\mathcal{G}$~\cite{gallager2013stochastic}:
\begin{equation}\label{S_eq:random_walk} 
\mathbf{p}_{r+1}= \mathbf{p}_r \, M
\end{equation}
where $\mathbf{p}_r$ is a $1 \times N$ vector with components denoting the probability of the random walk arriving at the respective node at discrete time $r$.

There are different continuous-time processes associated with the random walk~\cite{delvenne2013stability,  lambiotte2014random}. In particular, consider the rate matrix 
\begin{equation}\label{S_eq:QMatrix}
Q = M - I, 
\end{equation}
where $I_{N \times N}$ is the identity matrix. %
Note that $L=-Q$ is the \emph{random walk Laplacian}. 
We then define the continuous-time Markov process with semi-group $P(r)$ governed by the forward Kolmogorov equation
\begin{equation}\label{ForwardKolmogorov}
    \frac{dP}{dr} = P\:Q,
\end{equation}
which has the solution
\begin{equation}\label{S_eq:MarkovSemigroup}
    P(r) = e^{rQ}.
\end{equation}
which, under broad assumptions, converges to a unique stationary distribution $\boldsymbol{\pi}$, defined by
\begin{equation} \label{S_eq:stationary_dist_MC}
\boldsymbol{\pi} = \boldsymbol{\pi} \, M ,
\end{equation}
where $\boldsymbol{\pi}$ is a $1 \times N$ probability vector, which fulfills $\boldsymbol{\pi} L=0$~\cite{scheutzow2021convergence}. 

\subsubsection*{Markov Stability as a cost function for clustering algorithms}

The dynamics of the Markov chain with transition matrix $M$ defined on the nodes of the graph  can be exploited to get insights into the properties of the graph $G$ itself. 
Following~\cite{delvenne2010stability,lambiotte2014random}, each partition of the graph into $c$ communities corresponds to a $N \times c$ indicator matrix $H$ 
where $H_{ij} = 1$ if node $i$ is part of community $j$ and  $H_{ij} = 0$ otherwise. We can then define the clustered autocovariance matrix for partition $H$ as
\begin{equation}\label{S_eq:autocovariance_matrix}
    \mathcal{K}_r(H) = H^T \left[\Pi P(r) - \boldsymbol{\pi}^T \boldsymbol{\pi} \right] H.
\end{equation}
The diagonal elements $\mathcal{K}_r(H)_{ii}$ correspond to the probabilities that the Markov process starting in one community $i$ does not leave the community up to time $r$, whereas the off-diagonal elements correspond to the probabilities that the process has left the community in which it started by time $r$. It is important to remark that $r$ is an intrinsic time of the Markov process that is used to explore the graph structure and is clearly distinct from the physical time of some applications. To avoid confusion, it is customary in Markov Stability analysis to refer to $r$ and/or $s=\log_{10}(r)$ as the Markov \emph{scale}. Following these observations, we define the Markov Stability of a partition $H$ by
\begin{equation}\label{S_eq:MarkovStability}
    \mathcal{R}_r(H) = \min_{0\le l\le r} \text{Tr}\left[\mathcal{K}_l(H)\right]
    \approx \text{Tr}\left[\mathcal{K}_r(H)\right].
\end{equation}
The approximation in~\eqref{S_eq:MarkovStability} is supported by numerical simulations that suggest that $\text{Tr}\left[\mathcal{K}_r(H)\right]$ is monotonically decreasing in $r$.
The Markov Stability $\mathcal{R}_r(H)$ is thus a dynamical quality measure of the partition for each Markov scale $r$ which can be maximised to determine optimal partitions for a given graph and each scale of the associated Markov process. 

The objective is therefore to find a partition $H(s)$ that maximises Markov Stability up to a time horizon (scale) $s$ for the Markov process on the graph:
\begin{equation}
    \mathcal{R}_s\left(H(s)\right) = \max_H \mathcal{R}_s(H).
\end{equation}
Optimisation of Markov Stability for different Markov scales $s$ then leads to a series of partitions $H(s)$. %
For small Markov scales, the Markov process can only explore local neighbourhoods, which leads to a fine partition, whereas increasing the Markov scale widens the horizon of the Markov process so that larger areas of the graph are explored, which leads to coarser partitions~\cite{lambiotte2014random}. Hence, the notion of a community as detected by Markov Stability analysis is strictly based on the spread of a diffusion on the network.

\newpage

\section{Skill clusters at coarse resolution}
\label{sec:MS7}

One of the features of our multiscale analysis is the possibility of extracting clusterings at different levels of resolution that are inherently robust in the data. In our MS analysis (Fig.~\ref{fig: markov_stability}), we found a robust coarser partition (MS7) into seven skill clusters. Here we cover succinctly some of the findings for this skill clusters following the same procedure and format as for MS21 above. 

\begin{table}[H]
\centering
\caption{Summary of statistics for coarse resolution skill clusters (MS7).}
\label{tab:table_MS7}
\resizebox{0.99\textwidth}{!}{%
\begin{tabular}{lllllllll}
\hline
Index & Cluster  & \begin{tabular}[c]{@{}l@{}}No. of \\ Skills\end{tabular} & \begin{tabular}[c]{@{}l@{}}No. of \\ Mentions\end{tabular} & \begin{tabular}[c]{@{}l@{}}Average\\ Mention\end{tabular} & \begin{tabular}[c]{@{}l@{}}Semantic\\ Similarity\end{tabular} & \begin{tabular}[c]{@{}l@{}}Skill\\ Containment\end{tabular} & \begin{tabular}[c]{@{}l@{}}Closeness\\ Centrality\end{tabular} & \begin{tabular}[c]{@{}l@{}}Average\\ Salary\end{tabular} \\
\hline
1 & Business and Financial Management & 611 & 217631667 & 3.348 & 0.169 & 0.521 & 0.142 & 32600 \\
2 & Engineering and Operations Management & 1184 & 117181750 & 1.803 & 0.137 & 0.435 & 0.144 & 30680 \\
3 & Technical and Software Development & 816 & 113008925 & 1.739 & 0.145 & 0.577 & 0.130 & 39060 \\
4 & Sales and Marketing & 457 & 88738239 & 1.365 & 0.185 & 0.352 & 0.137 & 29310 \\
5 & Teaching and Healthcare & 688 & 56098105 & 0.863 & 0.139 & 0.356 & 0.134 & 29510 \\
6 & Hospitality and Food Industry & 85 & 18441250 & 0.284 & 0.179 & 0.193 & 0.137 & 22000 \\
7 & Information Security and Cybersecurity & 65 & 2903836 & 0.045 & 0.246 & 0.164 & 0.119 & 44110
\\
\hline
\end{tabular}%
}
\end{table}

\begin{figure}[h!]
    \centering
    \includegraphics[width = 0.85\textwidth, angle=0]{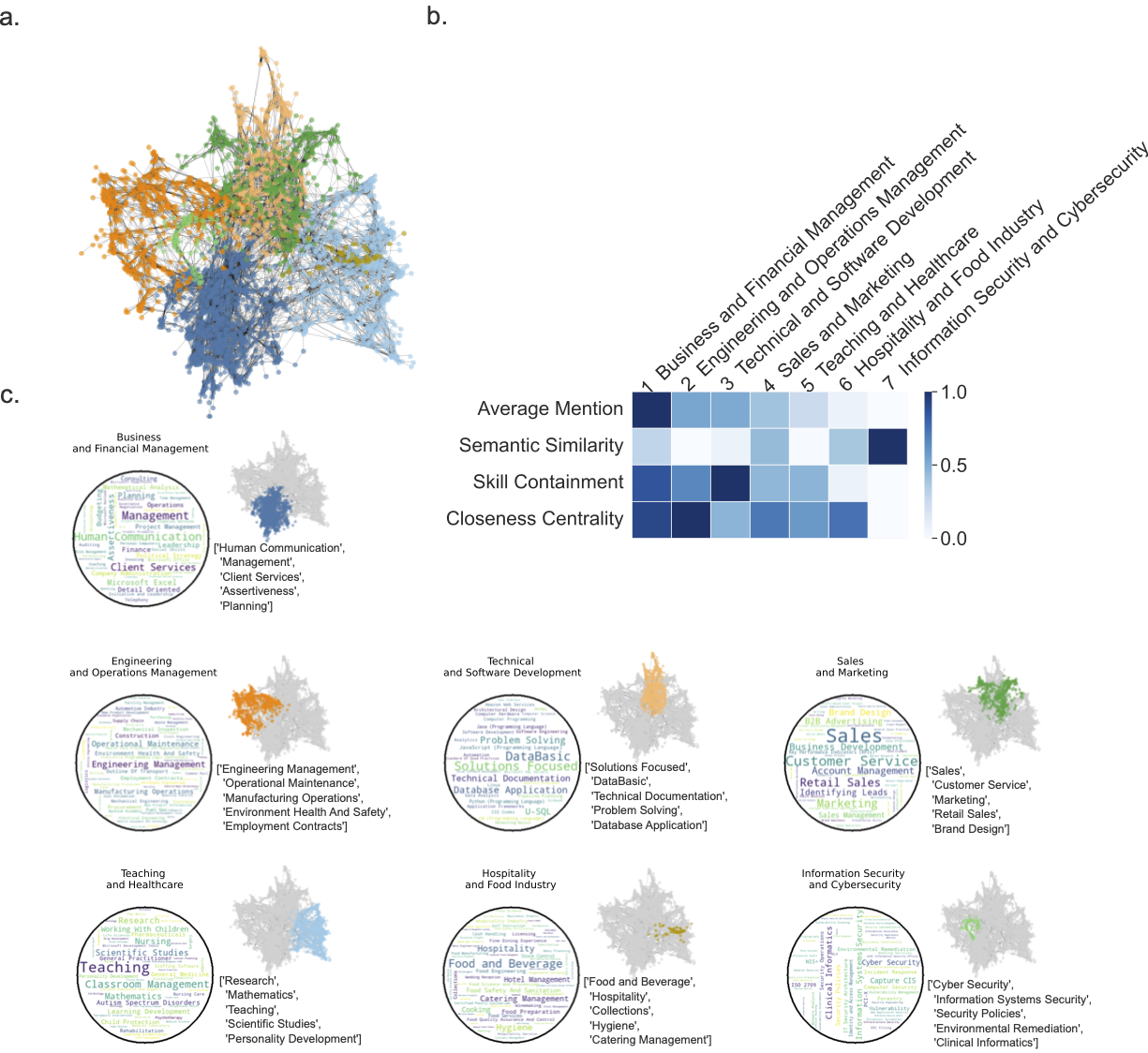}
     \caption{\textbf{Co-occurrence skill clusters (MS7)} (a) Skills network coloured according to the 7 skill clusters. (b) Summary heatmap of skill clusters properties. (c) For each of the 7 clusters, word cloud where font size represents skill eigenvector centrality, and list of top 5 most frequent skills. }
    \label{fig: summary_plot_MS7}
\end{figure}

Table~\ref{tab:table_MS7} and Figure~\ref{fig: summary_plot_MS7} present a summary of the results with computed statistics for the clusters, summary labels, and  word clouds.

Unlike MS21, we see greater imbalance in the number of skills contained in each cluster, with a twenty-fold difference between the cluster with the fewest skills (`Information Security and Cybersecurity', 65 skills) and the most skills (`Engineering and Operations Management', 1184 skills). Also, despite containing half as many skills, `Business and Financial Management' skills are mentioned twice as much (3.3 mentions per advert) than `Engineering and Operations Management' (1.8 mentions per advert). As expected, the within-cluster semantic similarity is broadly lower than in the MS21 configuration, reflecting the larger size of these clusters, which combine skills more different from one another. The `Information Security and Cybersecurity', the smallest cluster, has higher semantic similarity than the others, with an ostensibly similar skills profile. Indeed, the Sankey diagram in Figure~\ref{fig: sankey_allconfig} shows that this cluster is persistent across a range of scales, suggesting these skills occupy a distinct, isolated region in the skills network. 

\begin{figure}[htb!]
    \centering
    \includegraphics[width = .85\textwidth, angle=0]{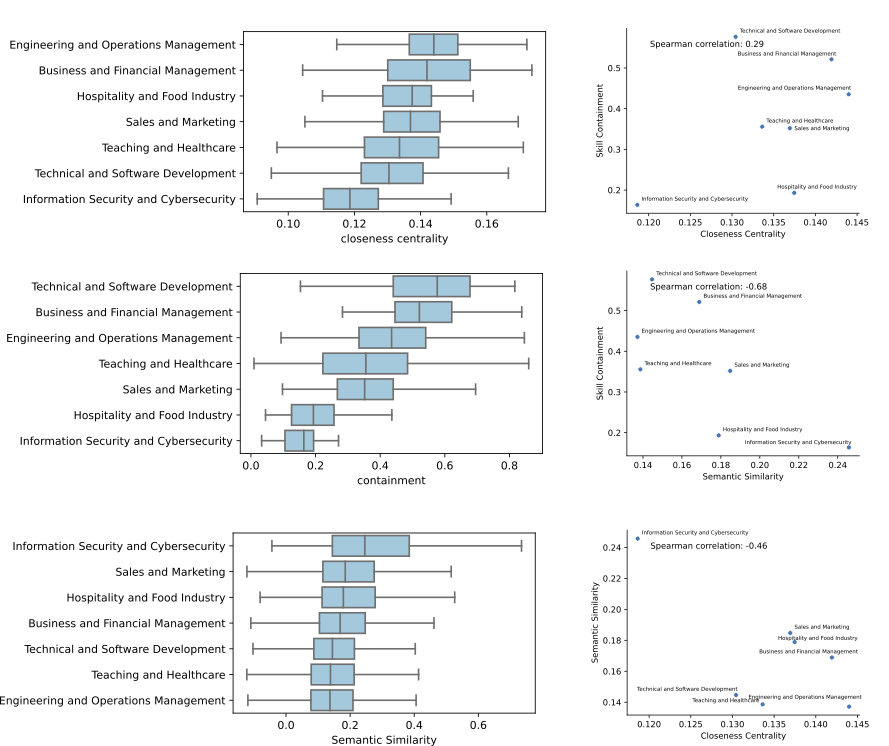}
     \caption{Boxplots for distributions of closeness centrality,  skill containment and within cluster semantic similarity for each MS7 skill cluster. The scatter plots compare these three variables.}
    \label{fig: box_plots_MS7}
\end{figure}

Compared to MS21, skill containment is generally higher, largely as a result of the MS7 clusters containing more skills. The highest containment is found in `Technical and Software Development', where 57.7\% of all co-occurrences between skills involve skills from the same cluster. Conversely, as in MS21, `Information Security and Cybersecurity' remains poorly contained (16.4\%) and is instead often connected to the rest of the skills network, yet with high semantic similarity and low closeness centrality. This reinforces the role of this skill cluster as a specialist cluster yet with broad co-occurrence. The differences in skill centrality across clusters also indicate the relevance of skills in specialised sectors, which are less shared across job adverts (i.e., low centrality) and with high containment, such as `Technical and Software Development'. See Figure~\ref{fig: box_plots_MS7} for more details of closeness centrality, skill containment and semantic similarity for the clusters in MS7.

This extensive overlap of skills into more sectorial groupings also leads to a convergence of the average salary, with 4 clusters having an average salary close to £30,000. The three outlying clusters are `Hospitality and Food industry' (£22,000) and `Technical and Software Development' (£39,060) and `Information Security and Cybersecurity' (£44,100), which can be viewed as specialist clusters with different levels of pay. 

\begin{figure}[htb!] 
    \centering
    \includegraphics[width = 1\textwidth]{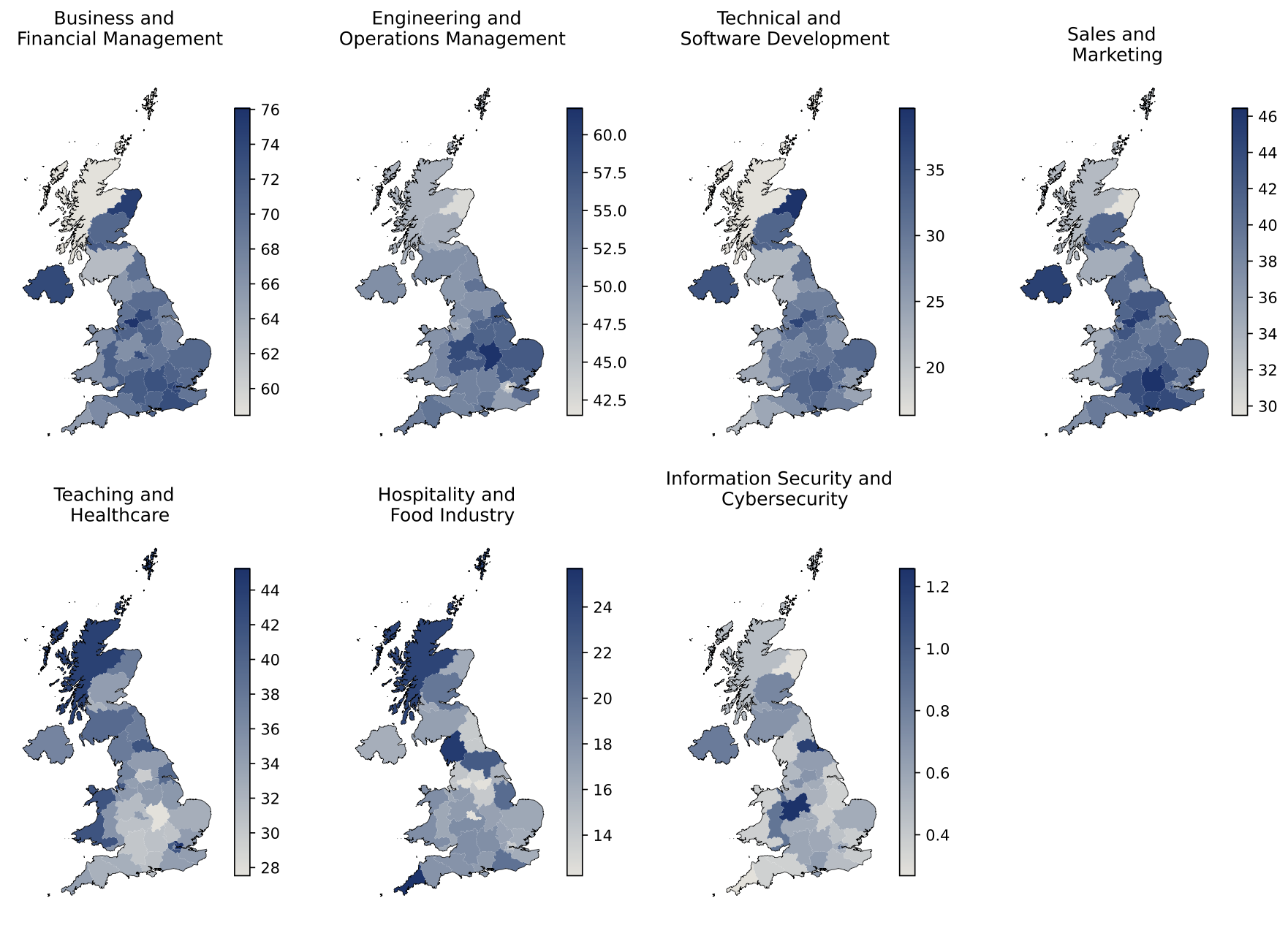}
     \caption{Maps for each of the MS7 clusters showing the percentage of all adverts in each NUTS2 regions featuring a skill from the cluster.}
    \label{fig: maps_MS7}
\end{figure}

Figure~\ref{fig: maps_MS7} presents the geographical distribution of the skill clusters in MS7 across NUTS2 regions. As for MS21, we observe substantial geographical variability that reflects different socio-economic factors, including industrial composition. This will be the object of future work.

\begin{figure}[htb!]
    \centering
    \includegraphics[width=.75\textwidth]{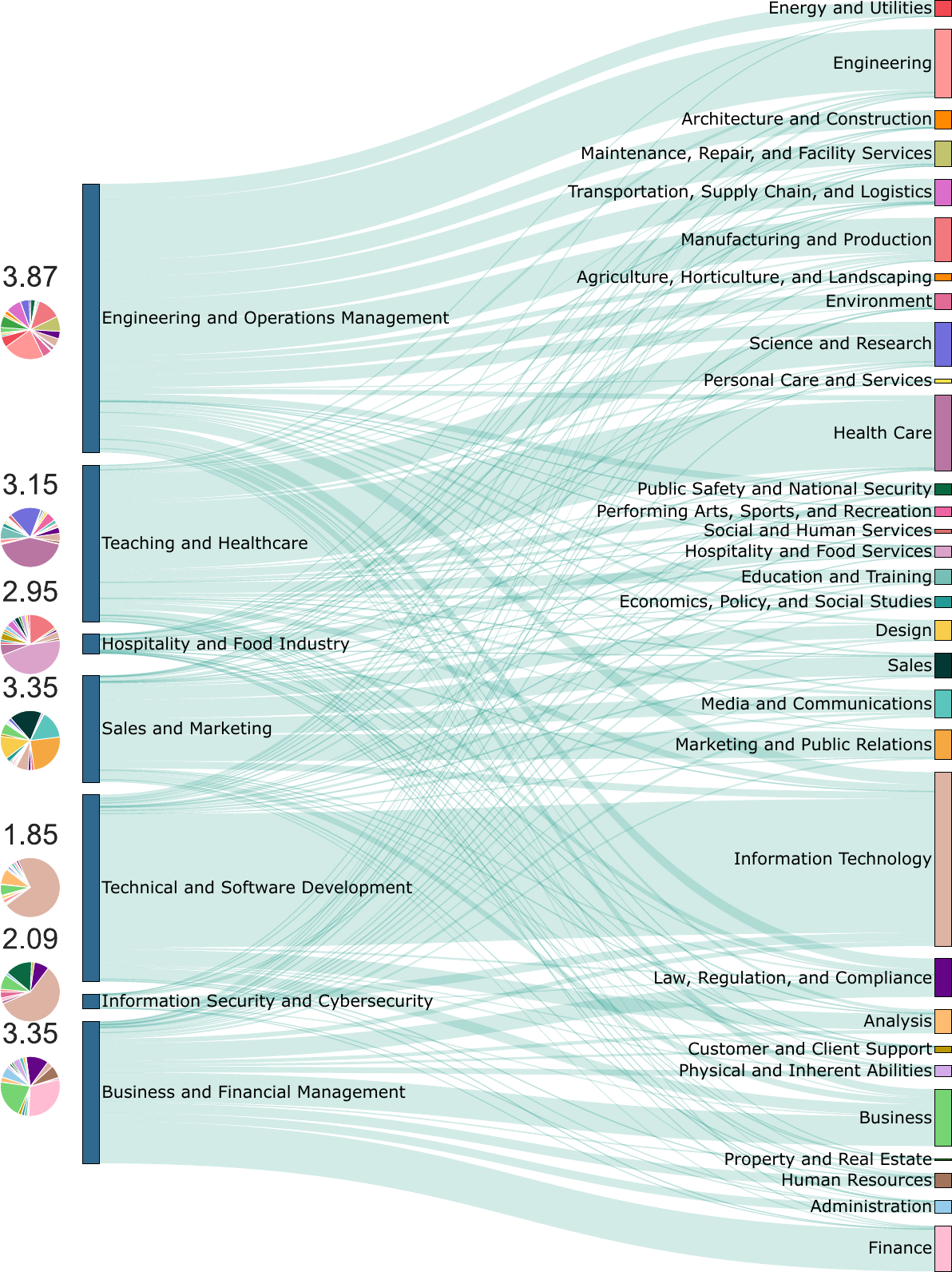}
    \caption{\textbf{Sankey diagram between co-ocurrence skill clusters (MS7) and expert-based skill clusters (Lightcast)}. There is broad agreement between the data-driven clusters and the LC categories in some skill areas where thematic content and co-occurrence match. For each cluster in MS7, we plot a pie chart to visualise the proportions of Lightcast categories, and the corresponding entropy to indicate how thematically mixed the cluster is.} 
    \label{fig: 7_lc_sankey}
\end{figure}

%
Finally,  Figure~\ref{fig: 7_lc_sankey} shows the Sankey diagram between the coarser MS7 skill clusters and the 32 LC skill categories. As for MS21, we find broad agreement, with some differences partly reflecting differences in the scale of the two categories (7 vs 32). For instance, `Technical and Software Development' in MS7 is closely related to the LC `Information Technology' category, but also incorporates the LC `Analysis' category. The LC `Health Care' category is included almost entirely within the MS7 `Teaching and Healthcare' cluster. Similarly, the MS7 `Sales and Marketing' cluster maps to the LC `Design', `Sales', `Media and Communications' and `Marketing and Public Relations' clusters. 
Note also that, unsurprisingly, as the MS clusters become coarser, the overall semantic similarity decreases since more dissimilar skills are grouped together (MS21: 0.172, MS7: 0.169, MS4: 0.141).

\newpage
\section{Code availability}
The code to generate the co-occurrence matrix from the JSON files can be accessed here: \url{https://github.com/Timliuzhaolu/Skills_clustering.git}.

\end{document}